\documentclass[aps,prl,a4,twocolumn,superscriptaddress,nofootinbib,showpacs,showkeys,preprintnumbers]{revtex4-1}
\pdfoutput=1
\usepackage{ulem}
\usepackage{color}
\usepackage[table]{xcolor}
\usepackage[colorlinks=true,citecolor=blue,linkcolor=blue,
breaklinks=true]{hyperref}
\usepackage{amsmath,amssymb}
\usepackage{graphicx}
\usepackage{slashed}
\usepackage{url}
\usepackage{makecell}
\usepackage{enumitem}
\usepackage{multirow}
\usepackage{array}
\usepackage{lipsum}
\usepackage{tabulary}
\usepackage{float}
\usepackage{booktabs}

\newcolumntype{?}{!{\vrule width 1pt}}
\newcolumntype{M}[1]{>{\centering\arraybackslash}m{#1}}
\newcommand{\lag}{\mathcal L}

\def\mET{\slashed{E}_T}

\begin{document}
%\preprint{HRI-RECAPP-2021-009}
%%%%%%%%%%%%%%%%%%%%%%%%%  Title  %%%%%%%%%%%%%%%%%%%%%%%%%%
%\title{An emergent $Z^\prime$ from the Higgs shadow}
\title{Emergent new symmetry from the Higgs shadow}

%%%%%%%%%%%%%%%%%%%%%%%%%  Authors  %%%%%%%%%%%%%%%%%%%%%%%%
\author{Waleed Abdallah}
\email{awaleed@sci.cu.edu.eg} 
%\affiliation{Regional Centre for Accelerator-based Particle Physics, %\\ 
%Harish-Chandra Research Institute, HBNI, \\
%Chhatnag Road, Jhunsi, Prayagraj (Allahabad) 211\,019, India}
\affiliation{Department of Mathematics, Faculty of Science, Cairo University, Giza 12613, Egypt}

\author{Anjan Kumar Barik}
\email{anjanbarik@hri.res.in}
%\affiliation{Regional Centre for Accelerator-based Particle Physics, %\\ 
%Harish-Chandra Research Institute, HBNI, \\
%Chhatnag Road, Jhunsi, Prayagraj (Allahabad) 211\,019, India}

\author{Santosh Kumar Rai}
\email{skrai@hri.res.in}
%\affiliation{Regional Centre for Accelerator-based Particle Physics, %\\ 
%Harish-Chandra Research Institute, HBNI, \\
%Chhatnag Road, Jhunsi, Prayagraj (Allahabad) 211\,019, India}

\author{Tousik Samui}
\email{tousiksamui@hri.res.in}
\affiliation{Regional Centre for Accelerator-based Particle Physics, %\\ 
Harish-Chandra Research Institute, HBNI, \\
Chhatnag Road, Jhunsi, Prayagraj (Allahabad) 211\,019, India}

%%%%%%%%%%%%%%%%%%%%%%%%%  Abstract  %%%%%%%%%%%%%%%%%%%%%%%
\begin{abstract}
We show in this Letter how a new hidden gauge symmetry responsible for neutrino mass as well as dark matter (DM)
in the Universe can be discovered through scalar mediators responsible for breaking the new symmetry. The new force 
mediator ($Z'$) may be lighter than the Standard Model~(SM) gauge bosons but cannot be observed in traditional 
searches for new gauge bosons. We highlight a novel way of discovering such a symmetry at the Large Hadron Collider (LHC) 
by incorporating an existing ATLAS analysis on four lepton final states which include the Higgs resonance. In addition, 
we show that the hidden sector also introduces flavor violation in the lepton sector which can become a significant channel of 
discovery for the new force. 
\end{abstract}

\maketitle
%%%%%%%%%%%%%%%%%%%%%%%%%  Main Body  %%%%%%%%%%%%%%%%%%%%%%
While the new century welcomed the discovery of a 125~GeV scalar~\cite{Aad:2012tfa,Chatrchyan:2012ufa} 
which completes the Standard Model~(SM) picture of observed particle spectrum, it also opened up the realm of the 
unknown structure beyond the SM. Despite the remarkable success of SM, several unexplained observations from 
experiments, be it neutrino mass or the existence of dark matter~(DM), have always hinted at the possibility of new 
physics beyond the SM~(BSM). However the sole discovery of the Higgs boson and nothing 
else at the Large Hadron Collider (LHC) has cast a shadow on 
what that exact possibility may be.  The non-observation of any new physics signal could be 
due to the presence  of very weakly interacting particles (defined by a symmetry that remains hidden in the LHC data) 
which may emerge in channels yet to be looked at by the experiments. In this Letter, we highlight that such a new symmetry 
may be lurking under the shadow of the most important discovery of the current century in particle physics, {\it viz.} the 
Higgs boson. We also show that this symmetry can provide  a solution to both the neutrino mass problem as well as DM signals. 

The importance of $U(1)$ symmetries have been significantly emphasized in new BSM ideas \cite{Kang:2004ix,Ma:1995xk,Barger:2004bz,deCarlos:1997yv,Ham:2007wc,Cvetic:1997ky,Langacker:2000ju,Langacker:2007ac,Kang:2009rd}
spanning neutrino physics, cosmology, DM, extra dimensions, and  supersymmetry, to name a few. 
They not only provide remedies to several outstanding issues in model 
building, but also manifest in our understanding of nature via several conserved global 
symmetries~\cite{Langacker:2008yv, Accomando:2013ita}. The hope of observing such a symmetry in 
experiments has seen a continuous effort over the last few decades at energy scales spanning a few MeV to several 
TeV. 
%\change{A discovery of $Z'$ and its decay could therefore lead us to an understanding of the underlying 
%quantum numbers the particles carry which could give hints about the underlying BSM physics.}{}
Our proposal of a hidden symmetry tries to present a novelty to the natural extension of the SM 
by a $U(1)$ gauge symmetry, by connecting the visible world with the {\it missing} (neutrinos and DM) while 
simultaneously suggesting why new physics still remains elusive to us at experiments like the LHC. 
We propose a neutrinophilic $U(1)_X$ extension~\cite{Abdallah:2021npg,Berbig:2020wve} to the SM
where any direct signal of the new symmetry is dependent on its overlap with SM particles. 
%%%NEW%%%
We have studied this model~\cite{Abdallah:2021npg}  in an earlier work featuring multi-lepton signals at the LHC 
through heavy neutrino production. A crucial part in that analysis was played by the non-vanishing gauge 
kinetic mixing (GKM) of the SM $U(1)_Y$ with the new $U(1)_X$ symmetry. 
%%%
In the absence of any meaningful mixing of $Z'$ with the $Z$ boson,
%between them, including the more obvious gauge kinetic mixing (GKM) of the SM $U(1)_Y$ with the new $U(1)_X$ symmetry, 
we highlight an interesting signal that would be able to probe the new $Z'$ directly via Higgs production 
and provide a robust signal in the current LHC run. 
As the $U(1)_X$ symmetry is broken by a singlet scalar, its admixture in the observed scalar at the LHC allows the 
$Z'$ to couple with the SM-like Higgs boson.
A similar production mechanism in the context of $U(1)_{B-L}$ for $Z'$ was considered in 
Refs.~\cite{Accomando:2017qcs, Amrith:2018yfb}. 
However, to obtain a light $Z'$ in that model and to avoid LHC constraints, the gauge coupling $g_{B-L}$ is 
restricted to unnaturally small values. This makes the contribution coming from
the second scalar insignificant. In our case, the new gauge coupling ($g_x$) 
remains naturally large with $g_x \sim g_2 \sim g_1$, where $g_1, g_2$ are coupling strengths for $U(1)_Y$ and
$SU(2)_L$ gauge symmetries, respectively.  We find that this gives a larger production rate for the $Z'$ through 
an additional scalar.

An additional phenomenologically interesting scenario occurs when the GKM vanishes completely. Then the $Z'$ in the model 
can become dominantly  
leptophilic. This happens when the decay of the $Z'$ is driven by one-loop contributions over the tree-level mode. 
An interesting outcome of the one-loop driven decay of the $Z'$ is lepton flavor violations (LFV) that could 
lead to interesting signatures of the new symmetry. The lightest of the heavy SM singlet neutrino of the model can also 
contribute to the cold dark matter (DM), since the light $Z'$ provides a new channel for a sufficient amount of neutrino annihilation 
into $Z'$ pair. %due to the electroweak strength gauge coupling $g_x$.
%provided we choose a conducive set of values for a few parameters. 
Thus one can summarise several interesting possibilities in a common framework:
\begin{itemize}
\item A light sub-100 GeV $Z'$ signal at the LHC via Higgs production.
\item A compatible fermionic DM with the correct relic density ensured by the presence of the light $Z'$.
\item Neutrino mass generation via inverse-seesaw mechanism~\cite{Abdallah:2021npg}\footnote{We ignore the 
possibility of generating neutrino mass radiatively in this study.}.
\item Lepton flavor violating (LFV) signal at one-loop through $Z'$ decay and possible contribution to lepton anomalous 
magnetic moments.
\end{itemize}

%===========================================================
\begin{table}[t!]
\begin{center}
\begin{tabular}{|c|c|c|c|c|c|}
\hline %& & & & &\\[-3mm]
Fields  & $SU(3)_C$ & $SU(2)_L$ & $U(1)_Y$ & $U(1)_X$ & Spin \\[1mm]
\hline %& & & & &\\[-3mm]
$H_{1(2)}$  & 1 & 2 & $-1/2$ & 0 ($-\, 1$) & 0 \\ [1mm]
\hline %& & & & &\\[-3mm]
$S$ & 1 & 1 & 0 & $2$ &  0 \\ [1mm]
\hline %& & & & &\\[-3mm]
$N_{L/R}^i$ & 1 & 1 & 0 & $1$  & 1/2 \\ [1mm]
\hline %& & & & &\\[-3mm]
\end{tabular}
\end{center}
\caption{New scalars ($H_1, H_2, S$) and matter fields~($N_L^i, N_R^i, \, i=1,2,3$) and their charge assignments
under the SM gauge group and $U(1)_X$.}
\label{tab:charges}
\end{table}
%===========================================================
%%%%%%%%%%%%%%%%%%%%%%%%%%%%%%%%%%%%%%%%%%%%%%%%%%%%%%%%%%%%
The model is an extension of the SM with an extra $U(1)_X$ gauge group 
and four new fields, including two chiral sterile neutrinos~$N_L, \,N_R$ added for each generation, 
an additional Higgs doublet $H_2$ and a scalar singlet $S$, where all the new fields are charged under 
$U(1)_X$ while all SM particles are neutral. 
The charge assignments of the new particles along with the first Higgs doublet~($H_1$)  
are listed in Table~\ref{tab:charges}.
The new scalar doublet ensures that a Dirac mass term for the neutrinos, necessary for the
inverse-seesaw mechanism, is guaranteed. All the new fields are charged under 
$U(1)_X$ while all SM particles are neutral. 
The new charge-neutral fermions mix with the SM neutrinos after symmetry breaking. With the assigned charges, 
the new gauge invariant Lagrangian added to the SM is given by (neglecting the kinetic terms)
\begin{eqnarray}
&\lag_S& \supset  - \mu_1 H_1^\dagger H_1\! -\! \mu_2 H_2^\dagger H_2 \!-\! \mu_s S^\dagger S  
                                      \!+\! \{\mu_{12} H_1^\dagger H_2 + {\rm\, h.c.} \}  \nonumber \\
 &-&   \lambda_1 (H_1^\dagger H_1)^2 - \lambda_2 (H_2^\dagger H_2)^2  
             - \lambda_s (S^\dagger S)^2  - \lambda'_{12} \left|H_1^\dagger H_2\right|^2  \nonumber \\
  &-&   \lambda_{12} H_1^\dagger H_1 H_2^\dagger H_2  - \lambda_{1s} H_1^\dagger H_1 S^\dagger S 
 - \lambda_{2s} H_2^\dagger H_2 S^\dagger S,  \nonumber \\[0.1cm] 
&\lag_{Y}& \supset  - \{ Y_\nu\,\overline l_L H_2 N_R \!+\! Y_R S \overline N_R N_R^C \!+\! Y_L S \overline N_L N_L^C \!+\! {\rm\, h.c.}\},  \nonumber \\[0.1cm] 
&\lag_{M}& \supset  - \hat{M}_N\left(\overline{N}_L N_R + \overline{N}_R N_L \right).
\label{eqn:lag}
\end{eqnarray}
We refer the readers to Ref.~\cite{Abdallah:2021npg} for more details on the model and its parameters, 
which lead to the masses and mixings of the fermions, scalars and gauge bosons. 
Note that $\mu_{12}$ is the coefficient of a soft-breaking term\footnote{This term can be generated 
dynamically by adding new scalars to the model.}  which breaks the $U(1)_X$ symmetry explicitly. Such a term 
can have its origin in a larger symmetry sitting at a much higher energy scale \cite{Das:2017fjf}.

After spontaneous breaking of the electroweak~(EW) and $U(1)_X$ symmetries when
$H_{1/2}$ and $S$ acquire VEVs, we are left with three physical CP-even 
neutral Higgs bosons, a charged Higgs and a pseudoscalar Higgs. 
The CP-even scalar mass matrix  in the $(\rho_1 \,\, \rho_2 \,\, \rho_3)^T$ basis~\cite{Abdallah:2021npg} is
\begin{eqnarray}
\small
\!\!\!\!\!\!\!M_H^2 = \begin{pmatrix}
2\lambda_1 v_1^2 + \mu_{12}\frac{v_2}{v_1}& \Lambda_{12} &~& \lambda_{1s}\,v_1 v_s \\
\Lambda_{12} & 2\lambda_2 v_2^2 + \mu_{12}\frac{v_1}{v_2} &~& \lambda_{2s}\,v_2 v_s \\
\lambda_{1s}\,v_1 v_s & \lambda_{2s}\,v_2 v_s &~& 2\lambda_s v_s^2
\end{pmatrix}\!\!, 
\end{eqnarray}
where $\Lambda_{12}=(\lambda_{12}+\lambda'_{12})v_1 v_2 -\mu_{12}$ and $v_{1 (2)}$ is the 
vacuum expectation value (VEV) of $H_{1 (2)}$. The three CP-even mass eigenstates $(h_1, h_2$ and $h_3)$ are linear 
combinations of the flavor states
%\change{,  $h_i = Z_{ij}^h \,\, \rho_j$, where $Z_{ij}^h$ represents the mixing matrix for the CP-even states}{
($\rho_i$) via the mixing matrix $Z^h$, {\it i.e.} $h_i = Z_{ij}^h \,\, \rho_j$. 
We identify $h_1$ as the $125$~GeV Higgs boson 
observed at the experiments and treat $h_2$ as the singlet dominated scalar. The other scalars
(belonging primarily to the $H_2$ doublet) are taken to be degenerate in mass and very heavy ($>3$~TeV). 
We shall work in the limit $\tan\beta = v_2/v_1  < 10^{-3}$ such that the only relevant admixture in
$h_1$ is from the singlet scalar which is parametrised by the mixing component $Z_{13}^h$. 

A non-zero VEV to $H_2$ charged under $U(1)_X$ allows a mixing between $Z$ and $Z'$. 
Note that the presence of $H_2$ and its participation in the symmetry breaking mechanism is crucial 
in giving mass to light neutrinos and generate the correct mixings in the neutrino sector~\cite{Abdallah:2021npg}. 
We show in Eq.~(\ref{tanthetap}) the resulting $Z$-$Z'$ mixing angle as a function of the gauge couplings and 
VEVs~\cite{Abdallah:2021npg},
\begin{equation}\label{tanthetap}	
\tan2\theta' = \dfrac{2g_z \left(g'_x v^2 + 2g_x v_2^2\right)}{{g'_x}^2 v^2 + 4 g_x g'_x v_2^2 + 4 g_x^2(v_2^2+ 4 v_s^2) 
- g_z^2 v^2} \,,  
\end{equation}
where $v=\sqrt{v_1^2+v_2^2} \simeq 246$~GeV, $g_z=\sqrt{g_1^2+g_2^2}$ and $g'_x$ measures 
the strength of gauge kinetic mixing. A desirable and natural choice for $Z$-$Z'$ mixing angle
would be to choose it smaller, such that it does not modify the $Z$ boson couplings with the SM fields. 
Precision data from LEP experiments on the $Z$ boson properties put an upper bound of $ 10^{-3}$ on 
$\theta'$~\cite{ParticleDataGroup:2020ssz}.  
The small value for $\theta'$ consistent with the LEP constraints can be easily achieved with 
$\tan\beta < 10^{-3}$ and $g_x' \simeq 0$, avoiding the need to fine-tune  $g_x$ to any unnatural value. 
We must note that kinetic mixing is unavoidable and will be generated at one-loop since $H_2$ is charged 
under both $U(1)_Y$ and $U(1)_X$.  However in a UV complete set-up it is possible to cancel the 
one-loop contribution to the kinetic mixing by introducing additional fields without affecting the phenomenology 
of the model \cite{delAguila:1995rb}. Such cancellations can help in obtaining much smaller values of $g_x'$, and 
the small GKM will make the $Z'$ interact very weakly with all SM matter fields, making it practically invisible 
and very difficult to observe. This would be akin to the situation we 
faced in our search for the SM Higgs boson which coupled very weakly to the light fermions and made it very difficult 
to discover the (now observed) SM Higgs boson at LEP and Tevatron.  
The main highlight of this work is to show the discovery channels for this invisible 
mediator and the importance of the Higgs sector of the model in producing the otherwise hidden $Z'$. 
We therefore work in the limit of small GKM and $\tan\beta$ values which help in avoiding strong limits on the 
$Z'$ mass and the gauge coupling $g_x$ \cite{Abdallah:2021npg}, and leads to interesting signals for $Z'$ decay.

We now focus on the limits that may arise from the Higgs sector.  Note that the scalars belonging to the second Higgs doublet 
are very heavy and satisfy the constraints trivially.  The constraints on the remaining scalars and their mixing is 
established using the publicly available 
packages {\tt HiggsBounds}~\cite{Bechtle:2008jh,Bechtle:2011sb}  and  {\tt HiggsSignals}~\cite{Bechtle:2013xfa}.  
These tools check for theoretical constraints on the Higgs mass as well as exclusions using the observed signal 
strength for the Higgs boson from LHC experiments. The experimental values are compared with expected deviations 
that may arise in any extended scalar sector which modifies the Higgs 
composition and its couplings. This allows us to obtain values for parameters in the Lagrangian which would be compatible 
with the observed Higgs boson and its decay probabilities~\cite{ATLAS:2016neq,Sopczak:2020vrs} in our model. 
More details of the constraints on the model parameters is given in our earlier work \cite{Abdallah:2021npg}.

Another interesting observation that we must highlight in our minimal model is the possibility to accommodate a DM 
candidate. We have three singlet left and right-handed neutrinos in the model and one of them becomes the DM if 
we appropriately choose one of the Yukawa couplings $Y_{\nu_{ij}}$ (shown in Eq.~(\ref{eqn:lag})) to be 
very small. We find that a pair of heavy neutrinos ($\nu_4$ and $\nu_5$), degenerate in mass get a decay lifetime
larger than the age of the Universe for $Y_{\nu_{11}} \lesssim 10^{-27}$ \footnote{This provides an upper bound
on the coupling strength and could also be chosen zero.} and $Y_{\nu_{1j}}=Y_{\nu_{j1}}=0$. 
This choice however makes the DM coupling to any of the SM states very weak, leading to the unwanted scenario of 
an overabundant DM in the Universe. The viability of DM being a thermal relic with the correct relic density is re-enforced
if the particle spectrum has non-SM lighter states to which the DM couples strongly enough, such that it can annihilate 
into these. This motivates us to choose the singlet scalar and $Z'$ lighter than the fermionic DM  candidates ($\nu_4$ and $\nu_5$). 
The dominant annihilation channels for the DM then become $\nu_4\nu_5\to h_2 Z^\prime$ ($t$-channel via 
$\nu_4$ and $\nu_5$), $\nu_4\nu_4\to Z^\prime Z^\prime (h_2 \, h_2), \nu_5\nu_5\to Z^\prime Z^\prime (h_2 \, h_2)$, 
where $h_2$ is the singlet $S$ dominated scalar. Here all interactions proceed via the unsuppressed $g_x$ coupling 
strength. 
\begin{table}[t!]
\begin{tabular}{|c|c|c|c|c|c|c|c|c|}\hline
 ${\lambda_{1}}$ & $\lambda_{s}$ & $\lambda_{2}$ & $\lambda_{12}$ & $\lambda^\prime_{12}$ & $\lambda_{2s}$ & \makecell{$\mu_{12}$\\ (GeV$^2$)} & \makecell{$v_s$\\ (GeV)} & $\tan\beta$  \\
\hline
 0.12875 & 1.044  & 1.0 & 0.005 & 0.005 & 0.0 & $10^{3}$ & 100 & $10^{-4}$   \\ \hline 
\end{tabular}
\caption{Scalar sector parameters consistent with all experimental constraints, where $(m_{h_1},m_{h_2})\!=\!(125,144.5)$~GeV.}
\label{tab:scalar}
\vspace{-0.5cm}
\end{table}
%

%\simeq 144.5$~GeV 

The light $Z'$ and $h_2$ which facilitate the DM to be a thermal relic can be observed 
%in the presence of light $Z'$, which could also be observed 
at the LHC through the Higgs mediated channel. To show how such a signal can be observed at the LHC and 
the simultaneous possibility of the DM relic being satisfied, we fix the scalar sector by choosing $h_1$ to be SM-like with 
mass $m_{h_1} \simeq 125$~GeV and keep the singlet $S$ dominated $h_2$ mass $m_{h_2}\simeq 144.5$~GeV, 
while the others are very heavy.  We choose the fixed values in Table~\ref{tab:scalar} since any other choice will 
not give any new characteristic features and only 
affect the overall production rate (through $v_s$). The most relevant parameters in our analysis become the 
\begin{figure}[h!]
\includegraphics[height=0.28\textwidth,width=0.47\textwidth]{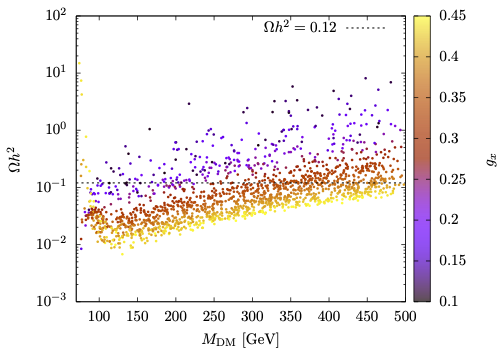} 
\vspace{-0.1cm}
\caption{The relic density $\Omega h^2$ as a function of the DM mass and the gauge coupling $g_x$.}
\label{Fig-DM-vPhM-LZp-1} 
\vspace{-0.15cm}
\end{figure}
%%%%%%%%%%%%%
gauge coupling $g_x$ that affects the $Z'$ mass and its coupling, and the quartic coupling $\lambda_{1s}$ which would 
affect the scalar sector mixing between the $H_1$ and $S$ components. We therefore scan over only these parameters 
and fix the other parameters of the model. Note that varying $\lambda_{1s}$ would also vary the $h_1$ and $h_2$ masses and 
is  constrained to be $< 0.033$ at the 3\,$\sigma$ level by
the measured Higgs boson mass of $125.25 \pm 0.17 $ GeV \cite{ParticleDataGroup:2020ssz}. This constrains the 
$h_2$ mass to be in the range 144.5-145 GeV.
%this variation is found to be less than a few percent as any large mixing is constrained by Higgs data. 
We show the DM relic density in Fig.~\ref{Fig-DM-vPhM-LZp-1} for the choice of parameter values and the scalar masses shown 
in Table~\ref{tab:scalar}. Quite clearly we can find a wide range of values for the DM in the model to satisfy the 
relic density requirements of $\Omega \, h^2 = 0.120 \pm 0.001$ at 90\% CL~\cite{Planck:2018vyg}.
The points are also allowed by {\tt XENON1T} and PandaX-4T~\cite{XENON:2018voc,XENON:2019rxp,PandaX-4T:2021bab} 
constraints on both spin-independent (SI) and spin-dependent (SD) direct detection cross sections. 
%\change{notwithstanding the fact that}{}
They also align with the indirect detection (annihilation of the DM pair to SM particles that could produce distinctive signatures in 
cosmic rays) constraints coming from the FERMI-LAT \cite{Daylan:2014rsa}, MAGIC \cite{MAGIC:2016xys} and 
PLANCK \cite{Planck:2018vyg} experiments
%are similarly satisfied 
in the given range due to the suppressed couplings with SM particles. The compatible points which give the correct 
relic density are found to have a typical DM-neucleon cross section of~$\lesssim 10^{-47}~{\rm cm}^2$.  We must 
however point out that the choice of parameters is not limited and the above is simply an example to show how easily 
a DM candidate could be accommodated in the model. As we show later, this can also play a crucial role for the LHC 
signal of $Z'$. A much more detailed DM analysis involving the full scan of the model parameter space is left for future work \cite{nuphilZp}.

\begin{figure}[h!]
\vspace{-0.1cm}
\includegraphics[width=0.50\textwidth,height=0.27\textwidth]{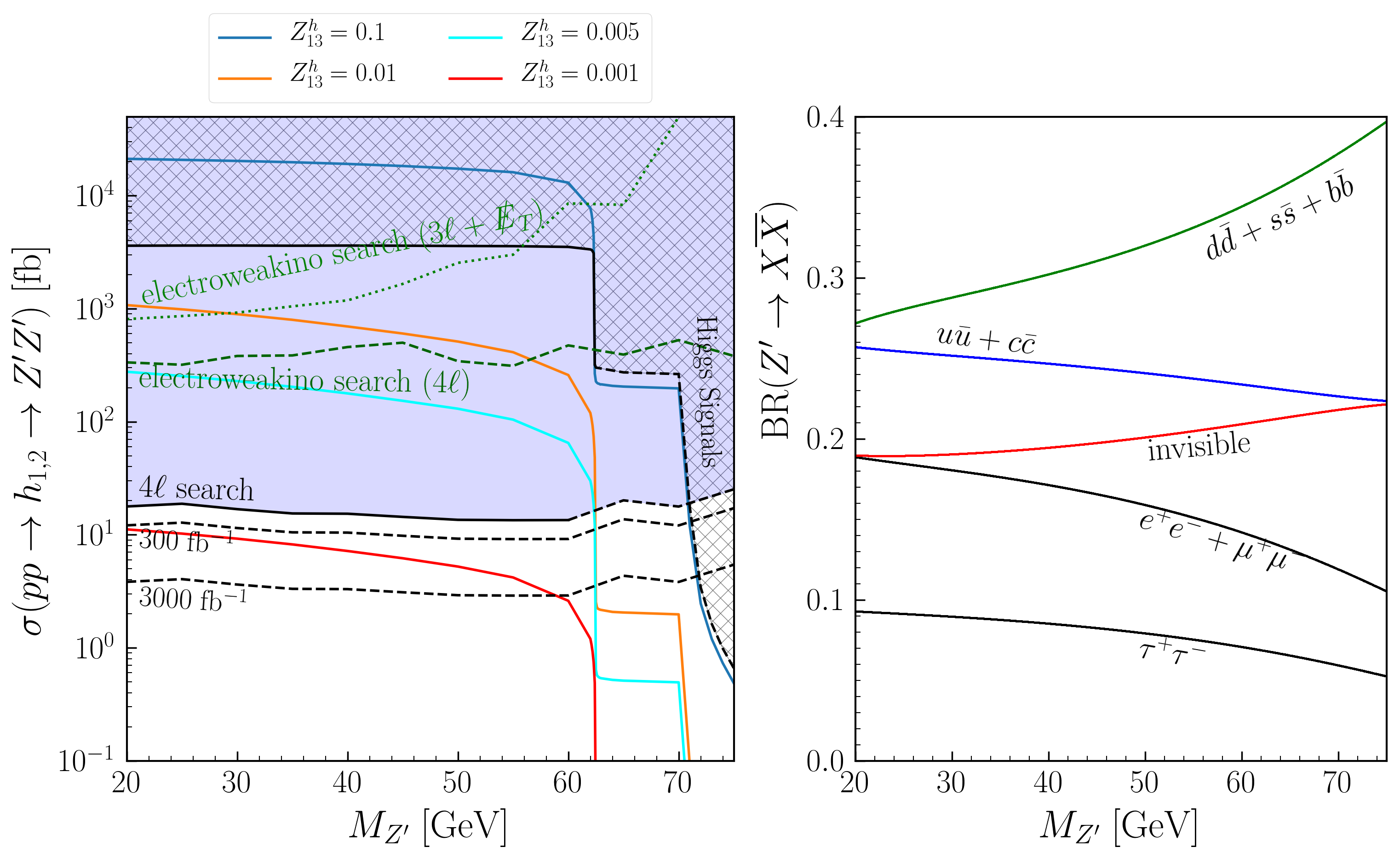}
\vspace{-0.5cm}
\caption{The pair production cross section of $Z'$ at the LHC via $h_1$ and $h_2$ and its tree-level decay branching ratios
along with current limits.} 
\label{fig:zpx}
\vspace{-0.3cm}
\end{figure}

We are now ready to analyse the light $Z'$ signal at the LHC based on the choice of parameters given in Table~\ref{tab:scalar}.
Note that the mixing $Z_{13}^h$ plays a crucial role in the pair production of $Z'$ at the LHC via  scalar mediators. 
%This mixing 
%dictates how much of the singlet admixture is required to have a substantial production cross section for a $Z'$ pair. 
%Together with the coupling of the $Z'$ to the singlet scalar with gauge coupling strength $g_x$, this channel 
%can give us a unique opportunity to study the otherwise weakly coupled $Z'$ through the scalar productions, via gluon-gluon fusion. 
With this mixing the scalar mediator can be produced via gluon-gluon fusion and will subsequently decay to
a  $Z'$ pair due to the gauge coupling $g_x$, giving us a unique opportunity to study the otherwise weakly coupled
$Z'$ boson. The value of the mixing parameter $Z_{13}^h$ determines the cross section of this process.
The $Z'$ would easily evade direct searches for very small values of GKM and $\tan\beta$. 
%($g_x' \lesssim 10^{-5}$ leads to a very small $\theta' \lesssim 10^{-5}$).  
As the SM particle couplings with $Z'$ are proportional to $\theta'$, 
the direct production of $Z'$ via fermion interaction is negligible for our chosen value of $g_x'$ and $\tan\beta$.
The scalar responsible for giving a mass to the $Z'$ therefore becomes the mediator which eventually helps in its 
production and its detection at experiments. We focus on the process 
\begin{align*}
p \, p \to h_{1,2} \to  Z' \, Z'  \, ,
\end{align*}
and show the $Z'$ pair production cross section at the LHC  in Fig.~\ref{fig:zpx} for different values of the mixing 
parameter $Z_{13}^h$. As expected, larger values of the mixing lead to significant production cross sections which 
could give clear hints of the $Z'$. The mixing $Z_{13}^h$ also affects the observed Higgs boson production and decay 
and is therefore constrained by Higgs boson measurements at LHC. %would however also affect the SM Higgs signal rates and 
%put constraints on its values. 
We can estimate the valid region of our signal cross section by scanning over the light 
$Z'$ mass ($70 \, {\rm GeV} \, > M_{Z'} > 20  \, {\rm GeV}$) and the
mixing parameter ($ 0.12 > |Z^h_{13}| > 0$) in the Higgs sector. The scan checks for the variation of Higgs signal strengths 
in the observed final states within allowed experimental bands at 95\% C.L. and also uses a requirement  on the 
allowed deviations in the Higgs branching to new modes to be less than 13\% (including invisible Higgs boson 
decays \cite{ATLAS:2018bnv}). This gives us a corresponding upper bound on $Z'$ pair production via the 
Higgs mediators, which is highlighted as the hatched region obtained using {\tt HiggsSignals}. The SM-like 
$h_1$ is the major contributor to $Z'$ production when $M_{Z'} \leq m_{h_1}/2$, while the singlet-dominated $h_2$ 
becomes the dominant contributor beyond this mass range. 
Note that the contribution from the singlet dominated $h_2$ is more or less constant for $M_{Z'} \leq m_{h_2}/2$ since 
the decay branching fraction of $h_2 \to Z' \, Z'$ is 100\%, a very specific feature of this model different from 
other $U(1)$ extensions considered in the literature. The off-shell contributions to the production fall rapidly for both 
$h_1$ and $h_2$ as is evident from Fig.~\ref{fig:zpx}. 
%This low production rate would become a region of interest for the high luminosity run of LHC. %\textcolor{red}{
In the region $ m_{h_2}/2 > M_{Z'} >62.5$ GeV, an increase in $Z'Z'$ production is only achieved through an increase in
$h_2$ production and, correspondingly, an increase in $Z^h_{13}$. This parameter is constrained by the observed 
signal strengths of the 125 GeV Higgs boson, so the total cross section in this region is also constrained to be 
small (see the hashed {\tt HiggsSignal} zone in Fig.~\ref{fig:zpx}).

%only way of increasing the on-shell $Z'Z'$ production will be through the enhancement of $h_2$ production.  This would imply going to larger values of $Z^h_{13}$ which are already constrained by the observed signal strengths of the 125 GeV Higgs boson. This happensbecause the large singlet scalar admixture in the SM-like Higgs affect its SM decay modes which in turn modify the Higgs signal strengths. This results into the {\tt HiggsSignal} hashed zone being spread out beyond $M_{Z'} >62.5$ for much lower 
%cross sections in this region. %} 
We however underline the importance of the above production channels as these might be the only 
relevant modes of observation of a light $Z'$ characterizing a {\it hidden gauge symmetry}. 

%%%
%The challenge in production does not always manifest itself in the decay properties of such a particle
The constraints on the production cross section do not significantly affect
the
decay, as the sub-100 GeV $Z'$ has a small total decay width,
$\Gamma_{Z'} \sim \mathcal{O}(10^{-9}-10^{-8})$~GeV but decays promptly (within the detector).
The branching ratios for $Z'$ decay are shown in Fig.~\ref{fig:zpx}. Using 
the parameter space of Fig.~\ref{fig:zpx} as representative points for the model we 
highlight the prospect of observing the $Z'$ signal at the LHC in its most sensitive $4\ell$ channel, 
where $\ell = e, \, \mu$. Although the $4\ell$ channel is the most likely channel of observing $Z'$ signal, 
other final states comprising of leptons and jets could also manifest as the channels of discovery. 
%We use {\tt CheckMATE}~\cite{Drees:2013wra} to confirm how much of the region of interest is still allowed by experiments 
%that have looked at leptonic final states with jets and missing transverse momenta. 
\begin{table*}[ht!]
	%	\centering
	%		\resizebox{8.5cm}{!}{
	\begin{tabular}{|c|c|c|}
		\cline{1-3}
		%	\multicolumn{1}{c|}{} 
		%                           & \multicolumn{4}{c}{Tree-level (One-loop) } 
		%                                                                          &  \multicolumn{4}{c|}{One-loop} 
		%                                                                                        &  \\  \hline	
		%   $Y_{\nu_{ij}} $ 
		Search Channel
		& Signal Region &   Kinematic Selection (GeV)  \\ \hline
		%                                                 & $ \sum \nu_a \nu_b$ & $\sum \bar{\ell}_{a} \ell_a$ & $\sum \bar{\ell}_{a} \ell_b$  & $\sum  j\,j$  

	$3\ell + \mET$	& {\tt A01} & $p_{T_{e(\mu)_{1}}} > 25(20)$,\,$p_{T_{e(\mu)_{2}}} > 15(10)$,\,$m_{\ell^{-}\ell^{+}} < 75$,\,$M_T < 100$,\,$\mET:(50-100)$  \\ \hline 
		
	$4\ell$	&({\tt G01},{\tt G02},{\tt G03}) &$p_{T_{e(\mu)_{1}}} > 25(20)$,\,$p_{T_{e(\mu)_{2}}} > 15(10)$,\,$m_{\ell^{-}\ell^{+}} < 75$,\,$ \mET:(0-50),(50-100),(100-150)$  \\ \hline                                                                                                                                         
	\end{tabular}
	%	}
	\caption{Kinematic selection and various signal regions in the $3\ell + \mET$ and $4\ell$ channels in Electroweakino searches at {\tt CMS}~\cite{CMS:2017moi}. The subleading leptons must satisfy $p_{T_{e(\mu)}} > 15 (10)$ GeV. If the leading lepton is a muon and the other leptons are electrons, the muon threshold is increased to $p_{T} > 25$ GeV.  Signal acceptance for the 
	$3\ell + \mET\, (4\ell)$ channel is around 5.3\% (17\%) for $M_Z'=25$ GeV and around 2.6\% (29\%) for $M_Z'=50$ GeV, respectively.
	}
	\label{tab:elwkino}
\end{table*}	
This implies that a multi-channel final state can constrain the parameter space of the model. Many searches extending  
over different multi-channel final states have been carried out by experiments at LHC in the context of different BSM models.
It is therefore natural to put our parameter region to test by determining the expected signal
yields in
search channels constrained by
the LHC experiments.
% into similar search channels and check its validity.
 A popular 
public code called {\tt CheckMATE}~\cite{Drees:2013wra} is used by the high-energy community to obtain limits on new BSM models from such experimental searches.  We obtain a weak bound from {\tt CheckMATE} which arises 
mainly from the multi-lepton searches for supersymmetric electroweakinos~\cite{CMS:2017moi}. %\textcolor{red} {
In the aforementioned search, different signal regions are identified based on the supersymmetric mass spectrum.  
The basic selection criteria employed for jets and leptons are: $p_{T_j} > 25$ GeV and $p_{T_{\ell}} > 10$ GeV with 
the same rapidity coverage of $|\eta| < 2.4$. The leptons and jets are isolated by $\Delta R_{j\ell} > 0.4$ and events with 
at least one b-jet are vetoed~\cite{CMS:2017moi}.
The dominant channel contributing to a bound for our parameter space comes from the 4$\ell$ final state as the $Z'$ can 
decay to electrons and muons combined with a probability of $18\%$ to $20\%$. The 4$\ell$ signal regions in 
Ref.~\cite{CMS:2017moi} are categorised using kinematic windows of $\mET$ as {\tt G01}, {\tt G02} and {\tt G03}. 
Their kinematic selection is listed in Table \ref{tab:elwkino} and the relevant constraint is shown in Fig.~\ref{fig:zpx} as 
{\it electroweakino search ($4\ell$)}. 
Note that the $Z'$ in our study is light and therefore the jets and leptons from its decay will have  low $p_T$. 
Hence some of the soft leptons may not satisfy the selection criteria. 
%The three lepton signal in our study will arise when a lepton does not satisfy the selection criteria in $Z' Z' \rightarrow 4 \ell$ channel as well as in $Z' Z' \rightarrow 2j\, 2 \ell$ 
%case, where one of the jets is missed and the other is mis-tagged as a lepton. 
For a $Z'$ boson mass less than 50 GeV, 
%the effective cross section of 3 lepton scenario will increase as 
a larger fraction of events in $Z' Z' \rightarrow 4 \ell$ do not the satisfy the 
selection criteria $p_{T_\ell} > 10$ GeV for all leptons. We find that 
the signal region {\tt A01} in Ref.~\cite{CMS:2017moi} provides the strongest constraint in the $3\ell + \mET$ channel.
We expect this constraint to be a bit stronger for lower $M_{Z'}$ values, whereas above 50 GeV of $M_{Z'}$ the constraint 
from $3\ell + \mET$ will get weaker. %as seen in Fig.~\ref{fig:zpx}.  
Note that the signal in our model will not have large $\mET$. 
%So even though the effective cross section of $3\ell + \mET$ is larger than $4 \ell$ in the $M_{Z'} < 50 $ GeV region, 
So the exclusion cut of $\mET < 50$ GeV as defined in {\tt CMS} analysis on $3\ell + \mET$ reduces its sensitivity significantly, 
as seen in Fig.~\ref{fig:zpx}.  %}%Although a strong $p_T$ on the charged lepton 
%reduces the sensitivity, the signal strength for the $Z'$ pair coming from the scalars is still affected by this search. 
The above bounds come from the direct search and are expectedly stronger than the {\tt HiggsSignal} bounds for $M_{Z'}<62.5$ GeV, as shown by the 
corresponding {\tt electroweakino search} exclusion curves in Fig.~\ref{fig:zpx}.  

The most relevant and strongest bound however comes from a $4\ell$ signal which was recently looked at 
by ATLAS~\cite{ATLAS:2021kog}. The analysis gives a differential cross section measurement of the $4$ lepton final 
state in the SM. This also happens to be the discovery channel for our $Z'$ and one expects to see resonant bumps 
in the dilepton invariant mass distributions. We can therefore use this analysis directly to test for hints of a $Z'$ boson. 
The ATLAS analysis has been included in the {\tt Rivet-3.1.4}~\cite{Buckley:2010ar,Bierlich:2019rhm} package, 
allowing a direct comparison of the experimental result with predictions of our signal. We include our model 
output for the aforementioned final state and use the package {\tt Contur}~\cite{Buckley:2021neu} to evaluate robust 
limits on the parameter space shown in Fig.~\ref{fig:zpx}. 
To put the bounds in perspective we note that in the $h_1 \to 4\ell$ signal region applicable 
to $20~{\rm GeV} \leq M_{Z'} \leq 62.5$~GeV, %\textcolor{red}{
the dilepton invariant mass variable $m_{34}$\footnote{Subscripts represent the $p_T$ ordered lepton numbers.} (Fig 7b) in Ref.~\cite{ATLAS:2021kog} has a larger 
bin wise SM cross section than $m_{12}$ (Fig 6b) in the $0-60$ GeV bin.  This is expected as the second 
pair of leptons arise from an off-shell $Z$ in the SM background events.  Although $m_{12}$ has a coarser bin in 
$0-50$ GeV,  most of the SM events have values of this variable near the $Z$ peak, and the total cross  section 
in the $0-50$ GeV bin is  $\mathcal{O}(10^{-2})$ fb. Hence $m_{12}$  constrains our parameter space 
most because our model results in a $Z'$ peak  in this region.  
For $M_{Z'}>62.5$ GeV the relevant measurement region is $m_{h_1}/2 < m_{4\ell} < m_{h_2}/2$~\cite{ATLAS:2021kog} 
in our analysis,  
%which is defined as off-shell $ZZ$ in Ref.~\cite{ATLAS:2021kog}, 
where the SM predicts the first and the second lepton pairs to come from on-shell and off-shell $Z$ bosons, respectively. 
Hence the second lepton pair 
invariant mass $m_{34}$ distribution gives a  tail above $60$ GeV, whereas for $m_{12}$ the cross section above 
$60$ GeV is higher compared to $m_{34}$ due to the presence of the SM $Z$ peak. In our model the relevant $Z'$ 
boson mass range is $75>M_{Z'}>62.5$~GeV and we expect a stronger constraint from 
$m_{34}$ where the SM background yields are smaller.    
The other kinematic variables including the full $4\ell$ invariant mass give weaker limits as the bin wise background 
cross section is much higher than in the relevant bins of the dilepton invariant mass, but the $4\ell$ invariant mass 
contributes in the final {\it chi-square} fit of {\tt contur}.  Note the specific bumps in the exclusion plots at $M_{Z'} \simeq 62, \, 72$~GeV where the sensitivity decreases 
suddenly. These points correspond to kinematic thresholds where $h_{1,2} \to Z' \, Z'$ become off-shell, leading to drop 
in signal yields. The measurement strategy in the ATLAS analysis employs  $50$  distributions 
in kinematic variables which can be also used to propose search sensitivity for our hypothesized $Z'$ boson for given values of $Z^h_{13}$ 
for high integrated luminosity options of the LHC. We show in Fig. \ref{fig:zpx}, the sensitivity curves with 300 fb$^{-1}$ and 
3000 fb$^{-1}$ integrated luminosity at the LHC, assuming a rather pessimistic view that similar efficiencies of the 
13~TeV analysis could be applicable to the 14~TeV run. %
%We have performed our analysis by implementing the full model in {\tt SARAH}~\cite{Staub:2013tta}. 
The events for the analysis were generated using  {\tt MadGraph5@aMCNLO}~\cite{Alwall:2011uj,Alwall:2014hca} 
and showered using {\tt Pythia \!8}~\cite{Sjostrand:2014zea}. 
The {\tt HepMC}~\cite{Dobbs:684090} output was then included in {\tt Rivet}~\cite{Buckley:2010ar,Bierlich:2019rhm} and 
run for the ATLAS analysis of Ref.~\cite{ATLAS:2021kog}.
The resulting {\tt YODA} file 
%that contains the recast information of our model 
was then input to {\tt Contur}~\cite{Butterworth:2016sqg,Buckley:2021neu} to evaluate a likelihood fit and determine the 
exclusions on the model parameter space. We propose that $h_1 \to 4\ell$ is the most sensitive channel to search for a light $Z'$ symbolising a hidden symmetry that couples weakly to SM particles 
and can lie hidden in the LHC data. 
%waiting to be discovered. 
%
\begin{table*}[t!]
%	\centering
%		\resizebox{8.5cm}{!}{
	\begin{tabular}{|c|c|c|c|c|c|}
		\cline{1-6}
%	\multicolumn{1}{c|}{} 
  %                           & \multicolumn{4}{c}{Tree-level (One-loop) } 
%                                                                          &  \multicolumn{4}{c|}{One-loop} 
 %                                                                                        &  \\  \hline	
     %   $Y_{\nu_{ij}} $ 
  $Z'$ decay
                             & $ \sum \nu_a \nu_b $ & $\sum \ell^+_{a} \ell^-_a$ & $\sum \ell^{\pm}_{a} \ell^{\mp}_b$ &  $ \sum j\,j$ 
%                                                 & $ \sum \nu_a \nu_b$ & $\sum \bar{\ell}_{a} \ell_a$ & $\sum \bar{\ell}_{a} \ell_b$  & $\sum  j\,j$  
                                                                                          & \multicolumn{1}{c|}{$\Gamma_{Z'}$ (GeV)}  \\ \hline
                                                                                        	
         \makecell{$Y_{\nu_{ij}}^{i \neq j} = 0, \, Y_\nu^{ii} \simeq 10^{-3}$} 
                             & $0.23 \, (10^{-3})$ & $0.16 \, (10^{-4})$ & $0 \, (10^{-4})$ & $0.61 \, (0)$ 
 %                                              & $$ & $10^{-6}$ & $10^{-7}$ & $0$  
                                                                                          &  $2 \times 10^{-9}$ \\ \hline 
          \makecell{$Y_{\nu_{ij}}^{i \neq j} \simeq 0.32, \, Y_\nu^{ii} \simeq 0.8$} 
                             & $10^{-3} \, (0.50)$ & $10^{-3} \, (0.340)$ & $0 \, (0.16)$ & $10^{-3} \, (0)$  
%                                               & $0.8$  & $0.135$ & $0.065$ & $0$
                                                                                          &  $1.130 \times 10^{-6}$ \\ \hline                                                                                                                                         
	\end{tabular}
%	}
\caption{Decay probabilities of $Z'$ at tree-level (one-loop) for different $Y_{\nu_{ij}}$. Here $\ell_a = \mu, \, \tau$, and
$M_{Z'} = 60$ GeV, $g_x = 0.3$, $Y_\nu^{11}=0$ and $Y_\nu^{1j}=0$ with $\theta' \simeq 10^{-5}$. 
}
\label{tab:zloopdk}
\end{table*}	
%%%%%%

\begin{figure}[h!]
\includegraphics[width=0.40\textwidth,height=0.25\textwidth]{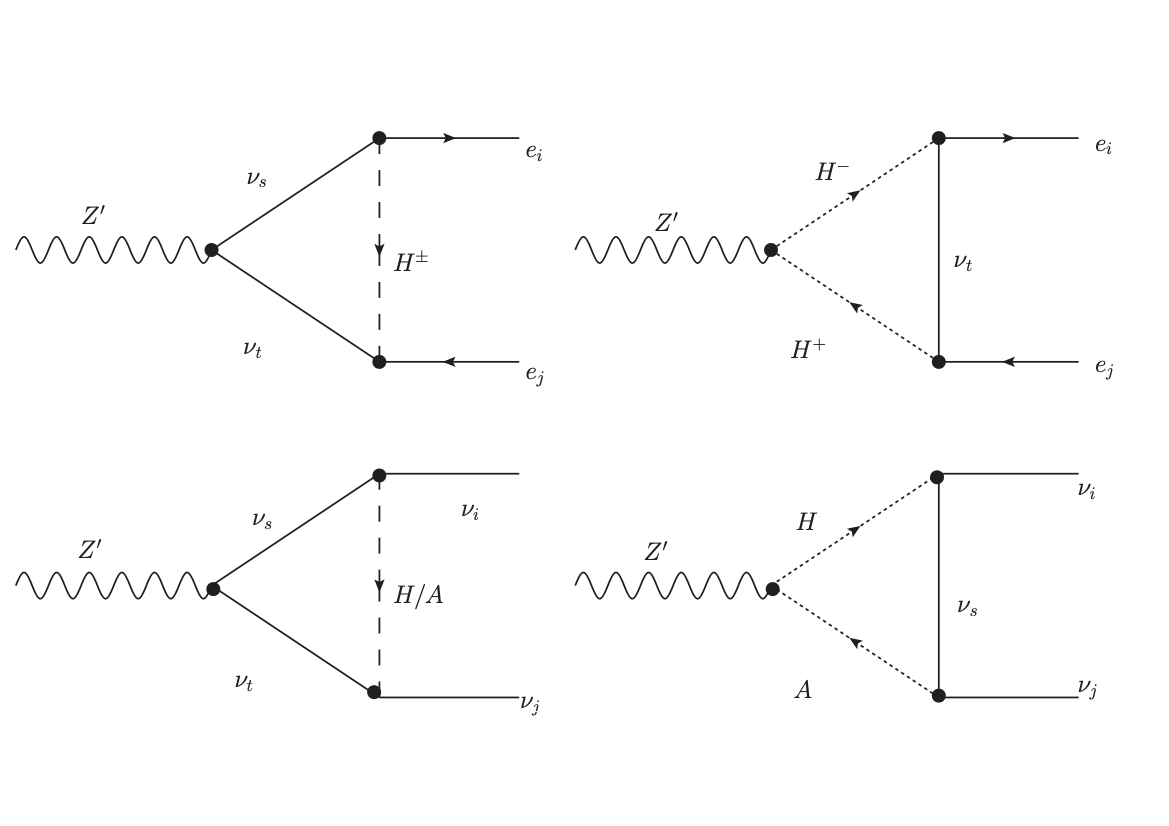}
\caption{Feynman diagrams for the one-loop decays of the $Z'$ boson to neutrinos and charged leptons. 
%The contributing diagrams have only the heavy neutrinos in the loop which couple with $g_x$ strength to $Z'$.
} 
\label{fig:zploopdk}
\end{figure}

\begin{figure}[h!]
\includegraphics[width=0.35\textwidth,height=0.26\textwidth]{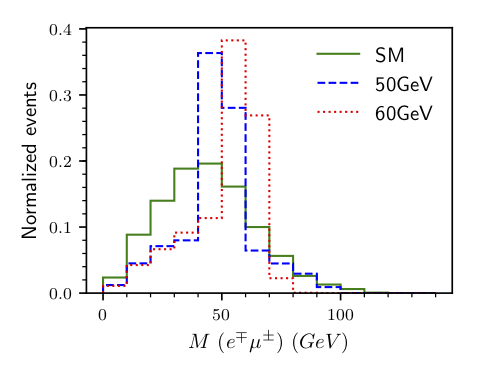}
\caption{Illustrating the invariant mass peaks in the wrong flavour mode, due to one-loop decay of the $Z'$ boson. 
%The contributing diagrams have only the heavy neutrinos in the loop which couple with $g_x$ strength to $Z'$.
} 
\label{fig:memu}
\end{figure}

We now comment on some interesting possibilities for our model which can have a major impact on the search strategies 
for the $Z'$ boson %. We note that when $Y_\nu \sim \mathcal{O}(0.1)$ 
when radiative decays of the $Z'$ boson become very important. 
%As $g_x \sim g_{_{\rm EW}}$ in our model while tree-level couplings with SM fermions is small due to the small $Z$-$Z'$ mixing, 
The term $Y_\nu\,\overline l_L H_2 N_R$ in the Lagrangian determines how large the radiative decay is. 
We find that for $Y_\nu \sim 10^{-1}$ the loop induced decays shown in Fig.~\ref{fig:zploopdk} have amplitudes which 
are proportional to $Y_\nu^2$ and start becoming comparable to the tree-level modes driven by $Z$-$Z'$ mixing. 
%The one-loop contributions to the decay width of $Z'$ are in the range 
%of $\sim \mathcal{O}(10^{-6}-10^{-4})$~GeV as compared to the tree-level width of $\mathcal{O}(10^{-9}-10^{-8})$~GeV 
%for $Y_\nu \sim 0.1$. 
An immediate and interesting consequence of this result is that the $Z'$ boson behaves as a leptophilic boson with no decay to quarks.
Note that the mixing of the light neutrino with heavy neutrinos is still very small unlike the typical inverse-seesaw mechanism, 
%the $\nu_{i}$-$Z$-$\nu_{j}$ coupling as well as $\nu_{j}$-$W$-$\ell$ coupling, where $i=1,2,3, \, j=4-9$, remains very small 
due to the choice of very small $\tan\beta$ values. 
%This ensures that the lepton unitarity and corrections to the $W$ boson mass are easily satisfied along with all existing limits 
%on production of heavy neutral leptons at experiments. 
%
We evaluate the one-loop decays of the $Z'$ boson and show a comparison of the decay branching ratios in 
Table~\ref{tab:zloopdk} for two choices of the Yukawa couplings.
The loop diagrams have been calculated analytically with the 
help of {\tt Package-X}~\cite{Patel:2015tea} and numerical results were obtained with the help 
of {\tt LoopTools}~\cite{Hahn:1998yk}.
The branching fraction of the $Z'$ boson to the exotic LFV modes could be arranged at the level of $15-25\%$ by suitably varying the $Y_\nu$ values. 
In a likely scenario of radiative decays dominating, the $Z'$ boson decays dominantly ($50\%$) to light neutrinos contributing 
to the invisible decay mode of the Higgs when $M_{Z'} \leq m_{h_1}/2$. The invisible mode would render the search for 
such a $Z'$ at the LHC 
very challenging. The most likely place of discovery in such a scenario would be  at future lepton colliders. 
%A clever tuning of 
%Note that the reduced branching ratio in the charged lepton mode would affect the $h_1 \to 4\ell$ signal, 
%relaxing the sensitivities obtained in Fig.~\ref{fig:zpx} but 
The leptophilic nature also leads to an increase in the charged lepton decay branching fraction of the $Z'$ boson to $50\%$, giving a  
stronger limit than what is obtained in Fig.~\ref{fig:zpx}. In addition, it opens up a more interesting possibility of observing 
LFV decays of the $Z'$ in the $h_1 \to 4\ell$ signal~\cite{Korner:1992an,Herrero:2018luu,Brdar:2020nbj}. The ATLAS analysis~\cite{ATLAS:2021kog} only looks at the invariant mass
distributions of opposite-sign-same-flavour leptons while in our case we will have a substantial decay mode of 
$Z' \to e\mu$ which will show up as an invariant mass peak in the wrong flavour mode (see Fig.~\ref{fig:memu}). 
We find that an improvement of up to 
60\% is achieved for signal sensitivity when this variable is included in the analysis. This would therefore provide a clear new
signal and huge improvement over the existing search for our model. A more unique possibility with very little SM background arises
if we consider only the LFV decay of $Z' \to e^\mp \mu^\pm$. This would give two pairs of same-flavor-same sign charged leptons. 
%As ($e^\mp e^\mp$ and $\mu^\pm \mu^\pm$) will have no invariant mass peaks, the analysis would rely on how well the charge measurement is carried out to tag these leptons. The only relevant backgrounds that may arise for SM will be from charge mis-measurements, which are well in control at LHC for the given $p_T$ range. 
%We however note that for $Z'$ mass less than 45 GeV, theSM background is large for $m_{e\mu}$ and therefore the signal sensitivity decrease compared  
%In addition, the possibility of lepton flavor violation (LFV) in $Z'$ via the loops can be studied at experiments, similar 
%to those considered in inverse-seesaw models to search for LFV in $Z$ decay~\cite{Korner:1992an,Herrero:2018luu,Brdar:2020nbj}. 

These LFV modes of the $Z'$ boson could then clearly be observed at the LHC by focusing on the LFV searches in the Higgs decay
which are of great current interest~\cite{CMS:2017con}.  A more promising search scenario would be in machines 
vouched as Higgs factories~\cite{Bambade:2019fyw, CEPCStudyGroup:2018ghi, FCC:2018evy, Calibbi:2021pyh} where 
the $h_1 \to Z' \, Z'$ mode can be looked at with more precision.  A close examination of the parameter space that could 
lead to such possibilities is interesting but beyond the scope of this work, and we leave it for future~studies.  

To conclude, we have highlighted through this Letter how a light $Z'$ in a popular and well motivated $U(1)$ extension of the 
SM can easily stay hidden in the LHC data if the new symmetry does not speak to the SM sector directly. We motivate 
such a scenario through the neutrino sector by introducing a neutrinophilic $U(1)$ which can also provide a DM candidate. 
The weakly interacting DM candidate gives the correct relic density when the $Z'$ boson is lighter than the DM candidate. We then highlight how such a light $Z'$ 
can be produced at the LHC via the Higgs channel. The existing LHC searches for such modes in the Higgs channel 
could be sensitive to a significant parameter space of our model which is explicitly shown by considering independent 
search channels and analyzing their effect on our parameter space. A very interesting consequence of the model is the 
decay of the $Z'$ boson via a loop when
it does not directly couple
with SM fields, giving the possibility of it being a leptophilic gauge
boson. The typically non-diagonal mixing in the light and heavy neutrino states and 
the structure of the $Y_\nu$ Yukawa coupling matrix can lead to LFV decays of $Z'$ which will open up more interesting 
signatures of the model at the LHC and future lepton colliders.  
%%%%%%%
\begin{acknowledgments}
The authors would like to acknowledge support from the Department of Atomic Energy, Government of India, for the 
Regional Centre for Accelerator-based Particle Physics~(RECAPP). 
 \end{acknowledgments}
%%%%%%
%\bibliography{nuphilZ}

\begin{thebibliography}{0}%
\makeatletter
\providecommand \@ifxundefined [1]{%
 \@ifx{#1\undefined}
}%
\providecommand \@ifnum [1]{%
 \ifnum #1\expandafter \@firstoftwo
 \else \expandafter \@secondoftwo
 \fi
}%
\providecommand \@ifx [1]{%
 \ifx #1\expandafter \@firstoftwo
 \else \expandafter \@secondoftwo
 \fi
}%
\providecommand \natexlab [1]{#1}%
\providecommand \enquote  [1]{``#1''}%
\providecommand \bibnamefont  [1]{#1}%
\providecommand \bibfnamefont [1]{#1}%
\providecommand \citenamefont [1]{#1}%
\providecommand \href@noop [0]{\@secondoftwo}%
\providecommand \href [0]{\begingroup \@sanitize@url \@href}%
\providecommand \@href[1]{\@@startlink{#1}\@@href}%
\providecommand \@@href[1]{\endgroup#1\@@endlink}%
\providecommand \@sanitize@url [0]{\catcode `\\12\catcode `\$12\catcode
  `\&12\catcode `\#12\catcode `\^12\catcode `\_12\catcode `\%12\relax}%
\providecommand \@@startlink[1]{}%
\providecommand \@@endlink[0]{}%
\providecommand \url  [0]{\begingroup\@sanitize@url \@url }%
\providecommand \@url [1]{\endgroup\@href {#1}{\urlprefix }}%
\providecommand \urlprefix  [0]{URL }%
\providecommand \Eprint [0]{\href }%
\providecommand \doibase [0]{http://dx.doi.org/}%
\providecommand \selectlanguage [0]{\@gobble}%
\providecommand \bibinfo  [0]{\@secondoftwo}%
\providecommand \bibfield  [0]{\@secondoftwo}%
\providecommand \translation [1]{[#1]}%
\providecommand \BibitemOpen [0]{}%
\providecommand \bibitemStop [0]{}%
\providecommand \bibitemNoStop [0]{.\EOS\space}%
\providecommand \EOS [0]{\spacefactor3000\relax}%
\providecommand \BibitemShut  [1]{\csname bibitem#1\endcsname}%
\let\auto@bib@innerbib\@empty
%</preamble>
\end{thebibliography}%


\begin{thebibliography}{53}%
\makeatletter
\providecommand \@ifxundefined [1]{%
 \@ifx{#1\undefined}
}%
\providecommand \@ifnum [1]{%
 \ifnum #1\expandafter \@firstoftwo
 \else \expandafter \@secondoftwo
 \fi
}%
\providecommand \@ifx [1]{%
 \ifx #1\expandafter \@firstoftwo
 \else \expandafter \@secondoftwo
 \fi
}%
\providecommand \natexlab [1]{#1}%
\providecommand \enquote  [1]{``#1''}%
\providecommand \bibnamefont  [1]{#1}%
\providecommand \bibfnamefont [1]{#1}%
\providecommand \citenamefont [1]{#1}%
\providecommand \href@noop [0]{\@secondoftwo}%
\providecommand \href [0]{\begingroup \@sanitize@url \@href}%
\providecommand \@href[1]{\@@startlink{#1}\@@href}%
\providecommand \@@href[1]{\endgroup#1\@@endlink}%
\providecommand \@sanitize@url [0]{\catcode `\\12\catcode `\$12\catcode
  `\&12\catcode `\#12\catcode `\^12\catcode `\_12\catcode `\%12\relax}%
\providecommand \@@startlink[1]{}%
\providecommand \@@endlink[0]{}%
\providecommand \url  [0]{\begingroup\@sanitize@url \@url }%
\providecommand \@url [1]{\endgroup\@href {#1}{\urlprefix }}%
\providecommand \urlprefix  [0]{URL }%
\providecommand \Eprint [0]{\href }%
\providecommand \doibase [0]{http://dx.doi.org/}%
\providecommand \selectlanguage [0]{\@gobble}%
\providecommand \bibinfo  [0]{\@secondoftwo}%
\providecommand \bibfield  [0]{\@secondoftwo}%
\providecommand \translation [1]{[#1]}%
\providecommand \BibitemOpen [0]{}%
\providecommand \bibitemStop [0]{}%
\providecommand \bibitemNoStop [0]{.\EOS\space}%
\providecommand \EOS [0]{\spacefactor3000\relax}%
\providecommand \BibitemShut  [1]{\csname bibitem#1\endcsname}%
\let\auto@bib@innerbib\@empty
%</preamble>
\bibitem [{\citenamefont {Aad}\ \emph {et~al.}(2012)\citenamefont {Aad} \emph
  {et~al.}}]{Aad:2012tfa}%
  \BibitemOpen
  \bibfield  {author} {\bibinfo {author} {\bibfnamefont {G.}~\bibnamefont
  {Aad}} \emph {et~al.} (\bibinfo {collaboration} {ATLAS}),\ }\href {\doibase
  10.1016/j.physletb.2012.08.020} {\bibfield  {journal} {\bibinfo  {journal}
  {Phys. Lett. B}\ }\textbf {\bibinfo {volume} {716}},\ \bibinfo {pages} {1}
  (\bibinfo {year} {2012})},\ \Eprint {http://arxiv.org/abs/1207.7214}
  {arXiv:1207.7214 [hep-ex]} \BibitemShut {NoStop}%
\bibitem [{\citenamefont {Chatrchyan}\ \emph {et~al.}(2012)\citenamefont
  {Chatrchyan} \emph {et~al.}}]{Chatrchyan:2012ufa}%
  \BibitemOpen
  \bibfield  {author} {\bibinfo {author} {\bibfnamefont {S.}~\bibnamefont
  {Chatrchyan}} \emph {et~al.} (\bibinfo {collaboration} {CMS}),\ }\href
  {\doibase 10.1016/j.physletb.2012.08.021} {\bibfield  {journal} {\bibinfo
  {journal} {Phys. Lett. B}\ }\textbf {\bibinfo {volume} {716}},\ \bibinfo
  {pages} {30} (\bibinfo {year} {2012})},\ \Eprint
  {http://arxiv.org/abs/1207.7235} {arXiv:1207.7235 [hep-ex]} \BibitemShut
  {NoStop}%
\bibitem [{\citenamefont {Kang}\ \emph {et~al.}(2005)\citenamefont {Kang},
  \citenamefont {Langacker},\ and\ \citenamefont {Li}}]{Kang:2004ix}%
  \BibitemOpen
  \bibfield  {author} {\bibinfo {author} {\bibfnamefont {J.-h.}\ \bibnamefont
  {Kang}}, \bibinfo {author} {\bibfnamefont {P.}~\bibnamefont {Langacker}}, \
  and\ \bibinfo {author} {\bibfnamefont {T.-j.}\ \bibnamefont {Li}},\ }\href
  {\doibase 10.1103/PhysRevD.71.015012} {\bibfield  {journal} {\bibinfo
  {journal} {Phys. Rev. D}\ }\textbf {\bibinfo {volume} {71}},\ \bibinfo
  {pages} {015012} (\bibinfo {year} {2005})},\ \Eprint
  {http://arxiv.org/abs/hep-ph/0411404} {arXiv:hep-ph/0411404} \BibitemShut
  {NoStop}%
\bibitem [{\citenamefont {Ma}(1996)}]{Ma:1995xk}%
  \BibitemOpen
  \bibfield  {author} {\bibinfo {author} {\bibfnamefont {E.}~\bibnamefont
  {Ma}},\ }\href {\doibase 10.1016/0370-2693(96)00524-2} {\bibfield  {journal}
  {\bibinfo  {journal} {Phys. Lett. B}\ }\textbf {\bibinfo {volume} {380}},\
  \bibinfo {pages} {286} (\bibinfo {year} {1996})},\ \Eprint
  {http://arxiv.org/abs/hep-ph/9507348} {arXiv:hep-ph/9507348} \BibitemShut
  {NoStop}%
\bibitem [{\citenamefont {Barger}\ \emph {et~al.}(2004)\citenamefont {Barger},
  \citenamefont {Kao}, \citenamefont {Langacker},\ and\ \citenamefont
  {Lee}}]{Barger:2004bz}%
  \BibitemOpen
  \bibfield  {author} {\bibinfo {author} {\bibfnamefont {V.}~\bibnamefont
  {Barger}}, \bibinfo {author} {\bibfnamefont {C.}~\bibnamefont {Kao}},
  \bibinfo {author} {\bibfnamefont {P.}~\bibnamefont {Langacker}}, \ and\
  \bibinfo {author} {\bibfnamefont {H.-S.}\ \bibnamefont {Lee}},\ }\href
  {\doibase 10.1016/j.physletb.2004.08.070} {\bibfield  {journal} {\bibinfo
  {journal} {Phys. Lett. B}\ }\textbf {\bibinfo {volume} {600}},\ \bibinfo
  {pages} {104} (\bibinfo {year} {2004})},\ \Eprint
  {http://arxiv.org/abs/hep-ph/0408120} {arXiv:hep-ph/0408120} \BibitemShut
  {NoStop}%
\bibitem [{\citenamefont {de~Carlos}\ and\ \citenamefont
  {Espinosa}(1997)}]{deCarlos:1997yv}%
  \BibitemOpen
  \bibfield  {author} {\bibinfo {author} {\bibfnamefont {B.}~\bibnamefont
  {de~Carlos}}\ and\ \bibinfo {author} {\bibfnamefont {J.~R.}\ \bibnamefont
  {Espinosa}},\ }\href {\doibase 10.1016/S0370-2693(97)00747-8} {\bibfield
  {journal} {\bibinfo  {journal} {Phys. Lett. B}\ }\textbf {\bibinfo {volume}
  {407}},\ \bibinfo {pages} {12} (\bibinfo {year} {1997})},\ \Eprint
  {http://arxiv.org/abs/hep-ph/9705315} {arXiv:hep-ph/9705315} \BibitemShut
  {NoStop}%
\bibitem [{\citenamefont {Ham}\ and\ \citenamefont {OH}(2007)}]{Ham:2007wc}%
  \BibitemOpen
  \bibfield  {author} {\bibinfo {author} {\bibfnamefont {S.~W.}\ \bibnamefont
  {Ham}}\ and\ \bibinfo {author} {\bibfnamefont {S.~K.}\ \bibnamefont {OH}},\
  }\href {\doibase 10.1103/PhysRevD.76.095018} {\bibfield  {journal} {\bibinfo
  {journal} {Phys. Rev. D}\ }\textbf {\bibinfo {volume} {76}},\ \bibinfo
  {pages} {095018} (\bibinfo {year} {2007})},\ \Eprint
  {http://arxiv.org/abs/0708.1785} {arXiv:0708.1785 [hep-ph]} \BibitemShut
  {NoStop}%
\bibitem [{\citenamefont {Cvetic}\ \emph {et~al.}(1997)\citenamefont {Cvetic},
  \citenamefont {Demir}, \citenamefont {Espinosa}, \citenamefont {Everett},\
  and\ \citenamefont {Langacker}}]{Cvetic:1997ky}%
  \BibitemOpen
  \bibfield  {author} {\bibinfo {author} {\bibfnamefont {M.}~\bibnamefont
  {Cvetic}}, \bibinfo {author} {\bibfnamefont {D.~A.}\ \bibnamefont {Demir}},
  \bibinfo {author} {\bibfnamefont {J.~R.}\ \bibnamefont {Espinosa}}, \bibinfo
  {author} {\bibfnamefont {L.~L.}\ \bibnamefont {Everett}}, \ and\ \bibinfo
  {author} {\bibfnamefont {P.}~\bibnamefont {Langacker}},\ }\href {\doibase
  10.1103/PhysRevD.56.2861} {\bibfield  {journal} {\bibinfo  {journal} {Phys.
  Rev. D}\ }\textbf {\bibinfo {volume} {56}},\ \bibinfo {pages} {2861}
  (\bibinfo {year} {1997})},\ \bibinfo {note} {[Erratum: Phys.Rev.D 58, 119905
  (1998)]},\ \Eprint {http://arxiv.org/abs/hep-ph/9703317}
  {arXiv:hep-ph/9703317} \BibitemShut {NoStop}%
\bibitem [{\citenamefont {Langacker}\ and\ \citenamefont
  {Plumacher}(2000)}]{Langacker:2000ju}%
  \BibitemOpen
  \bibfield  {author} {\bibinfo {author} {\bibfnamefont {P.}~\bibnamefont
  {Langacker}}\ and\ \bibinfo {author} {\bibfnamefont {M.}~\bibnamefont
  {Plumacher}},\ }\href {\doibase 10.1103/PhysRevD.62.013006} {\bibfield
  {journal} {\bibinfo  {journal} {Phys. Rev. D}\ }\textbf {\bibinfo {volume}
  {62}},\ \bibinfo {pages} {013006} (\bibinfo {year} {2000})},\ \Eprint
  {http://arxiv.org/abs/hep-ph/0001204} {arXiv:hep-ph/0001204} \BibitemShut
  {NoStop}%
\bibitem [{\citenamefont {Langacker}\ \emph {et~al.}(2008)\citenamefont
  {Langacker}, \citenamefont {Paz}, \citenamefont {Wang},\ and\ \citenamefont
  {Yavin}}]{Langacker:2007ac}%
  \BibitemOpen
  \bibfield  {author} {\bibinfo {author} {\bibfnamefont {P.}~\bibnamefont
  {Langacker}}, \bibinfo {author} {\bibfnamefont {G.}~\bibnamefont {Paz}},
  \bibinfo {author} {\bibfnamefont {L.-T.}\ \bibnamefont {Wang}}, \ and\
  \bibinfo {author} {\bibfnamefont {I.}~\bibnamefont {Yavin}},\ }\href
  {\doibase 10.1103/PhysRevLett.100.041802} {\bibfield  {journal} {\bibinfo
  {journal} {Phys. Rev. Lett.}\ }\textbf {\bibinfo {volume} {100}},\ \bibinfo
  {pages} {041802} (\bibinfo {year} {2008})},\ \Eprint
  {http://arxiv.org/abs/0710.1632} {arXiv:0710.1632 [hep-ph]} \BibitemShut
  {NoStop}%
\bibitem [{\citenamefont {Kang}\ \emph {et~al.}(2011)\citenamefont {Kang},
  \citenamefont {Langacker}, \citenamefont {Li},\ and\ \citenamefont
  {Liu}}]{Kang:2009rd}%
  \BibitemOpen
  \bibfield  {author} {\bibinfo {author} {\bibfnamefont {J.}~\bibnamefont
  {Kang}}, \bibinfo {author} {\bibfnamefont {P.}~\bibnamefont {Langacker}},
  \bibinfo {author} {\bibfnamefont {T.}~\bibnamefont {Li}}, \ and\ \bibinfo
  {author} {\bibfnamefont {T.}~\bibnamefont {Liu}},\ }\href {\doibase
  10.1007/JHEP04(2011)097} {\bibfield  {journal} {\bibinfo  {journal} {JHEP}\
  }\textbf {\bibinfo {volume} {04}},\ \bibinfo {pages} {097} (\bibinfo {year}
  {2011})},\ \Eprint {http://arxiv.org/abs/0911.2939} {arXiv:0911.2939
  [hep-ph]} \BibitemShut {NoStop}%
\bibitem [{\citenamefont {Langacker}(2009)}]{Langacker:2008yv}%
  \BibitemOpen
  \bibfield  {author} {\bibinfo {author} {\bibfnamefont {P.}~\bibnamefont
  {Langacker}},\ }\href {\doibase 10.1103/RevModPhys.81.1199} {\bibfield
  {journal} {\bibinfo  {journal} {Rev. Mod. Phys.}\ }\textbf {\bibinfo {volume}
  {81}},\ \bibinfo {pages} {1199} (\bibinfo {year} {2009})},\ \Eprint
  {http://arxiv.org/abs/0801.1345} {arXiv:0801.1345 [hep-ph]} \BibitemShut
  {NoStop}%
\bibitem [{\citenamefont {Accomando}\ \emph {et~al.}(2013)\citenamefont
  {Accomando}, \citenamefont {Becciolini}, \citenamefont {Belyaev},
  \citenamefont {De~Curtis}, \citenamefont {Dominici}, \citenamefont {King},
  \citenamefont {Moretti},\ and\ \citenamefont
  {Shepherd-Themistocleous}}]{Accomando:2013ita}%
  \BibitemOpen
  \bibfield  {author} {\bibinfo {author} {\bibfnamefont {E.}~\bibnamefont
  {Accomando}}, \bibinfo {author} {\bibfnamefont {D.}~\bibnamefont
  {Becciolini}}, \bibinfo {author} {\bibfnamefont {A.}~\bibnamefont {Belyaev}},
  \bibinfo {author} {\bibfnamefont {S.}~\bibnamefont {De~Curtis}}, \bibinfo
  {author} {\bibfnamefont {D.}~\bibnamefont {Dominici}}, \bibinfo {author}
  {\bibfnamefont {S.~F.}\ \bibnamefont {King}}, \bibinfo {author}
  {\bibfnamefont {S.}~\bibnamefont {Moretti}}, \ and\ \bibinfo {author}
  {\bibfnamefont {C.}~\bibnamefont {Shepherd-Themistocleous}},\ }\href
  {\doibase 10.22323/1.191.0125} {\bibfield  {journal} {\bibinfo  {journal}
  {PoS}\ }\textbf {\bibinfo {volume} {DIS2013}},\ \bibinfo {pages} {125}
  (\bibinfo {year} {2013})}\BibitemShut {NoStop}%
\bibitem [{\citenamefont {Abdallah}\ \emph
  {et~al.}(2021{\natexlab{a}})\citenamefont {Abdallah}, \citenamefont {Barik},
  \citenamefont {Rai},\ and\ \citenamefont {Samui}}]{Abdallah:2021npg}%
  \BibitemOpen
  \bibfield  {author} {\bibinfo {author} {\bibfnamefont {W.}~\bibnamefont
  {Abdallah}}, \bibinfo {author} {\bibfnamefont {A.~K.}\ \bibnamefont {Barik}},
  \bibinfo {author} {\bibfnamefont {S.~K.}\ \bibnamefont {Rai}}, \ and\
  \bibinfo {author} {\bibfnamefont {T.}~\bibnamefont {Samui}},\ }\href
  {\doibase 10.1103/PhysRevD.104.095031} {\bibfield  {journal} {\bibinfo
  {journal} {Phys. Rev. D}\ }\textbf {\bibinfo {volume} {104}},\ \bibinfo
  {pages} {095031} (\bibinfo {year} {2021}{\natexlab{a}})},\ \Eprint
  {http://arxiv.org/abs/2106.01362} {arXiv:2106.01362 [hep-ph]} \BibitemShut
  {NoStop}%
\bibitem [{\citenamefont {Berbig}\ \emph {et~al.}(2020)\citenamefont {Berbig},
  \citenamefont {Jana},\ and\ \citenamefont {Trautner}}]{Berbig:2020wve}%
  \BibitemOpen
  \bibfield  {author} {\bibinfo {author} {\bibfnamefont {M.}~\bibnamefont
  {Berbig}}, \bibinfo {author} {\bibfnamefont {S.}~\bibnamefont {Jana}}, \ and\
  \bibinfo {author} {\bibfnamefont {A.}~\bibnamefont {Trautner}},\ }\href
  {\doibase 10.1103/PhysRevD.102.115008} {\bibfield  {journal} {\bibinfo
  {journal} {Phys. Rev. D}\ }\textbf {\bibinfo {volume} {102}},\ \bibinfo
  {pages} {115008} (\bibinfo {year} {2020})},\ \Eprint
  {http://arxiv.org/abs/2004.13039} {arXiv:2004.13039 [hep-ph]} \BibitemShut
  {NoStop}%
\bibitem [{\citenamefont {Accomando}\ \emph {et~al.}(2018)\citenamefont
  {Accomando}, \citenamefont {Delle~Rose}, \citenamefont {Moretti},
  \citenamefont {Olaiya},\ and\ \citenamefont
  {Shepherd-Themistocleous}}]{Accomando:2017qcs}%
  \BibitemOpen
  \bibfield  {author} {\bibinfo {author} {\bibfnamefont {E.}~\bibnamefont
  {Accomando}}, \bibinfo {author} {\bibfnamefont {L.}~\bibnamefont
  {Delle~Rose}}, \bibinfo {author} {\bibfnamefont {S.}~\bibnamefont {Moretti}},
  \bibinfo {author} {\bibfnamefont {E.}~\bibnamefont {Olaiya}}, \ and\ \bibinfo
  {author} {\bibfnamefont {C.~H.}\ \bibnamefont {Shepherd-Themistocleous}},\
  }\href {\doibase 10.1007/JHEP02(2018)109} {\bibfield  {journal} {\bibinfo
  {journal} {JHEP}\ }\textbf {\bibinfo {volume} {02}},\ \bibinfo {pages} {109}
  (\bibinfo {year} {2018})},\ \Eprint {http://arxiv.org/abs/1708.03650}
  {arXiv:1708.03650 [hep-ph]} \BibitemShut {NoStop}%
\bibitem [{\citenamefont {Amrith}\ \emph {et~al.}(2019)\citenamefont {Amrith},
  \citenamefont {Butterworth}, \citenamefont {Deppisch}, \citenamefont {Liu},
  \citenamefont {Varma},\ and\ \citenamefont {Yallup}}]{Amrith:2018yfb}%
  \BibitemOpen
  \bibfield  {author} {\bibinfo {author} {\bibfnamefont {S.}~\bibnamefont
  {Amrith}}, \bibinfo {author} {\bibfnamefont {J.~M.}\ \bibnamefont
  {Butterworth}}, \bibinfo {author} {\bibfnamefont {F.~F.}\ \bibnamefont
  {Deppisch}}, \bibinfo {author} {\bibfnamefont {W.}~\bibnamefont {Liu}},
  \bibinfo {author} {\bibfnamefont {A.}~\bibnamefont {Varma}}, \ and\ \bibinfo
  {author} {\bibfnamefont {D.}~\bibnamefont {Yallup}},\ }\href {\doibase
  10.1007/JHEP05(2019)154} {\bibfield  {journal} {\bibinfo  {journal} {JHEP}\
  }\textbf {\bibinfo {volume} {05}},\ \bibinfo {pages} {154} (\bibinfo {year}
  {2019})},\ \Eprint {http://arxiv.org/abs/1811.11452} {arXiv:1811.11452
  [hep-ph]} \BibitemShut {NoStop}%
\bibitem [{\citenamefont {Das}\ \emph {et~al.}(2018)\citenamefont {Das},
  \citenamefont {Li}, \citenamefont {Nandi},\ and\ \citenamefont
  {Rai}}]{Das:2017fjf}%
  \BibitemOpen
  \bibfield  {author} {\bibinfo {author} {\bibfnamefont {K.}~\bibnamefont
  {Das}}, \bibinfo {author} {\bibfnamefont {T.}~\bibnamefont {Li}}, \bibinfo
  {author} {\bibfnamefont {S.}~\bibnamefont {Nandi}}, \ and\ \bibinfo {author}
  {\bibfnamefont {S.~K.}\ \bibnamefont {Rai}},\ }\href {\doibase
  10.1140/epjc/s10052-017-5495-0} {\bibfield  {journal} {\bibinfo  {journal}
  {Eur. Phys. J. C}\ }\textbf {\bibinfo {volume} {78}},\ \bibinfo {pages} {35}
  (\bibinfo {year} {2018})},\ \Eprint {http://arxiv.org/abs/1708.00328}
  {arXiv:1708.00328 [hep-ph]} \BibitemShut {NoStop}%
 \bibitem [{\citenamefont {Zyla}\ \emph {et~al.}(2020)\citenamefont {Zyla} \emph
  {et~al.}}]{ParticleDataGroup:2020ssz}%
  \BibitemOpen
  \bibfield  {author} {\bibinfo {author} {\bibfnamefont {P.~A.}\ \bibnamefont
  {Zyla}} \emph {et~al.} (\bibinfo {collaboration} {Particle Data Group}),\
  }\href {\doibase 10.1093/ptep/ptaa104} {\bibfield  {journal} {\bibinfo
  {journal} {PTEP}\ }\textbf {\bibinfo {volume} {2020}},\ \bibinfo {pages}
  {083C01} (\bibinfo {year} {2020})}\BibitemShut {NoStop}% 
\bibitem [{\citenamefont {del Aguila}\ \emph {et~al.}(1995)\citenamefont {del
  Aguila}, \citenamefont {Masip},\ and\ \citenamefont
  {Perez-Victoria}}]{delAguila:1995rb}%
  \BibitemOpen
  \bibfield  {author} {\bibinfo {author} {\bibfnamefont {F.}~\bibnamefont {del
  Aguila}}, \bibinfo {author} {\bibfnamefont {M.}~\bibnamefont {Masip}}, \ and\
  \bibinfo {author} {\bibfnamefont {M.}~\bibnamefont {Perez-Victoria}},\ }\href
  {\doibase 10.1016/0550-3213(95)00511-6} {\bibfield  {journal} {\bibinfo
  {journal} {Nucl. Phys. B}\ }\textbf {\bibinfo {volume} {456}},\ \bibinfo
  {pages} {531} (\bibinfo {year} {1995})},\ \Eprint
  {http://arxiv.org/abs/hep-ph/9507455} {arXiv:hep-ph/9507455} \BibitemShut
  {NoStop}%
\bibitem [{\citenamefont {Bechtle}\ \emph {et~al.}(2014)\citenamefont
  {Bechtle}, \citenamefont {Heinemeyer}, \citenamefont {St\r{a}l},
  \citenamefont {Stefaniak},\ and\ \citenamefont {Weiglein}}]{Bechtle:2013xfa}%
  \BibitemOpen
  \bibfield  {author} {\bibinfo {author} {\bibfnamefont {P.}~\bibnamefont
  {Bechtle}}, \bibinfo {author} {\bibfnamefont {S.}~\bibnamefont {Heinemeyer}},
  \bibinfo {author} {\bibfnamefont {O.}~\bibnamefont {St\r{a}l}}, \bibinfo
  {author} {\bibfnamefont {T.}~\bibnamefont {Stefaniak}}, \ and\ \bibinfo
  {author} {\bibfnamefont {G.}~\bibnamefont {Weiglein}},\ }\href {\doibase
  10.1140/epjc/s10052-013-2711-4} {\bibfield  {journal} {\bibinfo  {journal}
  {Eur. Phys. J. C}\ }\textbf {\bibinfo {volume} {74}},\ \bibinfo {pages}
  {2711} (\bibinfo {year} {2014})},\ \Eprint {http://arxiv.org/abs/1305.1933}
  {arXiv:1305.1933 [hep-ph]} \BibitemShut {NoStop}%
\bibitem [{\citenamefont {Bechtle}\ \emph {et~al.}(2010)\citenamefont
  {Bechtle}, \citenamefont {Brein}, \citenamefont {Heinemeyer}, \citenamefont
  {Weiglein},\ and\ \citenamefont {Williams}}]{Bechtle:2008jh}%
  \BibitemOpen
  \bibfield  {author} {\bibinfo {author} {\bibfnamefont {P.}~\bibnamefont
  {Bechtle}}, \bibinfo {author} {\bibfnamefont {O.}~\bibnamefont {Brein}},
  \bibinfo {author} {\bibfnamefont {S.}~\bibnamefont {Heinemeyer}}, \bibinfo
  {author} {\bibfnamefont {G.}~\bibnamefont {Weiglein}}, \ and\ \bibinfo
  {author} {\bibfnamefont {K.~E.}\ \bibnamefont {Williams}},\ }\href {\doibase
  10.1016/j.cpc.2009.09.003} {\bibfield  {journal} {\bibinfo  {journal}
  {Comput. Phys. Commun.}\ }\textbf {\bibinfo {volume} {181}},\ \bibinfo
  {pages} {138} (\bibinfo {year} {2010})},\ \Eprint
  {http://arxiv.org/abs/0811.4169} {arXiv:0811.4169 [hep-ph]} \BibitemShut
  {NoStop}%
\bibitem [{\citenamefont {Bechtle}\ \emph {et~al.}(2011)\citenamefont
  {Bechtle}, \citenamefont {Brein}, \citenamefont {Heinemeyer}, \citenamefont
  {Weiglein},\ and\ \citenamefont {Williams}}]{Bechtle:2011sb}%
  \BibitemOpen
  \bibfield  {author} {\bibinfo {author} {\bibfnamefont {P.}~\bibnamefont
  {Bechtle}}, \bibinfo {author} {\bibfnamefont {O.}~\bibnamefont {Brein}},
  \bibinfo {author} {\bibfnamefont {S.}~\bibnamefont {Heinemeyer}}, \bibinfo
  {author} {\bibfnamefont {G.}~\bibnamefont {Weiglein}}, \ and\ \bibinfo
  {author} {\bibfnamefont {K.~E.}\ \bibnamefont {Williams}},\ }\href {\doibase
  10.1016/j.cpc.2011.07.015} {\bibfield  {journal} {\bibinfo  {journal}
  {Comput. Phys. Commun.}\ }\textbf {\bibinfo {volume} {182}},\ \bibinfo
  {pages} {2605} (\bibinfo {year} {2011})},\ \Eprint
  {http://arxiv.org/abs/1102.1898} {arXiv:1102.1898 [hep-ph]} \BibitemShut
  {NoStop}%
\bibitem [{\citenamefont {Aad}\ \emph {et~al.}(2016)\citenamefont {Aad} \emph
  {et~al.}}]{ATLAS:2016neq}%
  \BibitemOpen
  \bibfield  {author} {\bibinfo {author} {\bibfnamefont {G.}~\bibnamefont
  {Aad}} \emph {et~al.} (\bibinfo {collaboration} {ATLAS, CMS}),\ }\href
  {\doibase 10.1007/JHEP08(2016)045} {\bibfield  {journal} {\bibinfo  {journal}
  {JHEP}\ }\textbf {\bibinfo {volume} {08}},\ \bibinfo {pages} {045} (\bibinfo
  {year} {2016})},\ \Eprint {http://arxiv.org/abs/1606.02266} {arXiv:1606.02266
  [hep-ex]} \BibitemShut {NoStop}%
\bibitem [{\citenamefont {Sopczak}(2020)}]{Sopczak:2020vrs}%
  \BibitemOpen
  \bibfield  {author} {\bibinfo {author} {\bibfnamefont {A.}~\bibnamefont
  {Sopczak}} (\bibinfo {collaboration} {ATLAS, CMS}),\ }\href {\doibase
  10.22323/1.353.0006} {\bibfield  {journal} {\bibinfo  {journal} {PoS}\
  }\textbf {\bibinfo {volume} {FFK2019}},\ \bibinfo {pages} {006} (\bibinfo
  {year} {2020})},\ \Eprint {http://arxiv.org/abs/2001.05927} {arXiv:2001.05927
  [hep-ex]} \BibitemShut {NoStop}%
\bibitem [{\citenamefont {Aghanim}\ \emph {et~al.}(2020)\citenamefont {Aghanim}
  \emph {et~al.}}]{Planck:2018vyg}%
  \BibitemOpen
  \bibfield  {author} {\bibinfo {author} {\bibfnamefont {N.}~\bibnamefont
  {Aghanim}} \emph {et~al.} (\bibinfo {collaboration} {Planck}),\ }\href
  {\doibase 10.1051/0004-6361/201833910} {\bibfield  {journal} {\bibinfo
  {journal} {Astron. Astrophys.}\ }\textbf {\bibinfo {volume} {641}},\ \bibinfo
  {pages} {A6} (\bibinfo {year} {2020})},\ \bibinfo {note} {[Erratum:
  Astron.Astrophys. 652, C4 (2021)]},\ \Eprint
  {http://arxiv.org/abs/1807.06209} {arXiv:1807.06209 [astro-ph.CO]}
  \BibitemShut {NoStop}%
\bibitem [{\citenamefont {Aprile}\ \emph {et~al.}(2018)\citenamefont {Aprile}
  \emph {et~al.}}]{XENON:2018voc}%
  \BibitemOpen
  \bibfield  {author} {\bibinfo {author} {\bibfnamefont {E.}~\bibnamefont
  {Aprile}} \emph {et~al.} (\bibinfo {collaboration} {XENON}),\ }\href
  {\doibase 10.1103/PhysRevLett.121.111302} {\bibfield  {journal} {\bibinfo
  {journal} {Phys. Rev. Lett.}\ }\textbf {\bibinfo {volume} {121}},\ \bibinfo
  {pages} {111302} (\bibinfo {year} {2018})},\ \Eprint
  {http://arxiv.org/abs/1805.12562} {arXiv:1805.12562 [astro-ph.CO]}
  \BibitemShut {NoStop}%
\bibitem [{\citenamefont {Aprile}\ \emph {et~al.}(2019)\citenamefont {Aprile}
  \emph {et~al.}}]{XENON:2019rxp}%
  \BibitemOpen
  \bibfield  {author} {\bibinfo {author} {\bibfnamefont {E.}~\bibnamefont
  {Aprile}} \emph {et~al.} (\bibinfo {collaboration} {XENON}),\ }\href
  {\doibase 10.1103/PhysRevLett.122.141301} {\bibfield  {journal} {\bibinfo
  {journal} {Phys. Rev. Lett.}\ }\textbf {\bibinfo {volume} {122}},\ \bibinfo
  {pages} {141301} (\bibinfo {year} {2019})},\ \Eprint
  {http://arxiv.org/abs/1902.03234} {arXiv:1902.03234 [astro-ph.CO]}
  \BibitemShut {NoStop}%
\bibitem [{\citenamefont {Meng}\ \emph {et~al.}(2021)\citenamefont {Meng} \emph
  {et~al.}}]{PandaX-4T:2021bab}%
  \BibitemOpen
  \bibfield  {author} {\bibinfo {author} {\bibfnamefont {Y.}~\bibnamefont
  {Meng}} \emph {et~al.} (\bibinfo {collaboration} {PandaX-4T}),\ }\href@noop
  {} {\  (\bibinfo {year} {2021})},\ \Eprint {http://arxiv.org/abs/2107.13438}
  {arXiv:2107.13438 [hep-ex]} \BibitemShut {NoStop}%
\bibitem [{\citenamefont {Daylan}\ \emph {et~al.}(2016)\citenamefont {Daylan},
  \citenamefont {Finkbeiner}, \citenamefont {Hooper}, \citenamefont {Linden},
  \citenamefont {Portillo}, \citenamefont {Rodd},\ and\ \citenamefont
  {Slatyer}}]{Daylan:2014rsa}%
  \BibitemOpen
  \bibfield  {author} {\bibinfo {author} {\bibfnamefont {T.}~\bibnamefont
  {Daylan}}, \bibinfo {author} {\bibfnamefont {D.~P.}\ \bibnamefont
  {Finkbeiner}}, \bibinfo {author} {\bibfnamefont {D.}~\bibnamefont {Hooper}},
  \bibinfo {author} {\bibfnamefont {T.}~\bibnamefont {Linden}}, \bibinfo
  {author} {\bibfnamefont {S.~K.~N.}\ \bibnamefont {Portillo}}, \bibinfo
  {author} {\bibfnamefont {N.~L.}\ \bibnamefont {Rodd}}, \ and\ \bibinfo
  {author} {\bibfnamefont {T.~R.}\ \bibnamefont {Slatyer}},\ }\href {\doibase
  10.1016/j.dark.2015.12.005} {\bibfield  {journal} {\bibinfo  {journal} {Phys.
  Dark Univ.}\ }\textbf {\bibinfo {volume} {12}},\ \bibinfo {pages} {1}
  (\bibinfo {year} {2016})},\ \Eprint {http://arxiv.org/abs/1402.6703}
  {arXiv:1402.6703 [astro-ph.HE]} \BibitemShut {NoStop}%
\bibitem [{\citenamefont {Ahnen}\ \emph {et~al.}(2016)\citenamefont {Ahnen}
  \emph {et~al.}}]{MAGIC:2016xys}%
  \BibitemOpen
  \bibfield  {author} {\bibinfo {author} {\bibfnamefont {M.~L.}\ \bibnamefont
  {Ahnen}} \emph {et~al.} (\bibinfo {collaboration} {MAGIC, Fermi-LAT}),\
  }\href {\doibase 10.1088/1475-7516/2016/02/039} {\bibfield  {journal}
  {\bibinfo  {journal} {JCAP}\ }\textbf {\bibinfo {volume} {02}},\ \bibinfo
  {pages} {039} (\bibinfo {year} {2016})},\ \Eprint
  {http://arxiv.org/abs/1601.06590} {arXiv:1601.06590 [astro-ph.HE]}
  \BibitemShut {NoStop}%
\bibitem [{\citenamefont {Abdallah}\ \emph
  {et~al.}(2021{\natexlab{b}})\citenamefont {Abdallah}, \citenamefont {Barik},
  \citenamefont {Rai},\ and\ \citenamefont {Samui}}]{nuphilZp}%
  \BibitemOpen
  \bibfield  {author} {\bibinfo {author} {\bibfnamefont {W.}~\bibnamefont
  {Abdallah}}, \bibinfo {author} {\bibfnamefont {A.~K.}\ \bibnamefont {Barik}},
  \bibinfo {author} {\bibfnamefont {S.~K.}\ \bibnamefont {Rai}}, \ and\
  \bibinfo {author} {\bibfnamefont {T.}~\bibnamefont {Samui}},\ }\href@noop {}
  {\bibfield  {journal} {\bibinfo  {journal} {Work in Progress}\ } (\bibinfo
  {year} {2021}{\natexlab{b}})}\BibitemShut {NoStop}%
\bibitem [{\citenamefont {Aaboud}\ \emph {et~al.}(2019)\citenamefont {Aaboud}
  \emph {et~al.}}]{ATLAS:2018bnv}%
  \BibitemOpen
  \bibfield  {author} {\bibinfo {author} {\bibfnamefont {M.}~\bibnamefont
  {Aaboud}} \emph {et~al.} (\bibinfo {collaboration} {ATLAS}),\ }\href
  {\doibase 10.1016/j.physletb.2019.04.024} {\bibfield  {journal} {\bibinfo
  {journal} {Phys. Lett. B}\ }\textbf {\bibinfo {volume} {793}},\ \bibinfo
  {pages} {499} (\bibinfo {year} {2019})},\ \Eprint
  {http://arxiv.org/abs/1809.06682} {arXiv:1809.06682 [hep-ex]} \BibitemShut
  {NoStop}%
\bibitem [{\citenamefont {Drees}\ \emph {et~al.}(2015)\citenamefont {Drees},
  \citenamefont {Dreiner}, \citenamefont {Schmeier}, \citenamefont
  {Tattersall},\ and\ \citenamefont {Kim}}]{Drees:2013wra}%
  \BibitemOpen
  \bibfield  {author} {\bibinfo {author} {\bibfnamefont {M.}~\bibnamefont
  {Drees}}, \bibinfo {author} {\bibfnamefont {H.}~\bibnamefont {Dreiner}},
  \bibinfo {author} {\bibfnamefont {D.}~\bibnamefont {Schmeier}}, \bibinfo
  {author} {\bibfnamefont {J.}~\bibnamefont {Tattersall}}, \ and\ \bibinfo
  {author} {\bibfnamefont {J.~S.}\ \bibnamefont {Kim}},\ }\href {\doibase
  10.1016/j.cpc.2014.10.018} {\bibfield  {journal} {\bibinfo  {journal}
  {Comput. Phys. Commun.}\ }\textbf {\bibinfo {volume} {187}},\ \bibinfo
  {pages} {227} (\bibinfo {year} {2015})},\ \Eprint
  {http://arxiv.org/abs/1312.2591} {arXiv:1312.2591 [hep-ph]} \BibitemShut
  {NoStop}%
\bibitem [{\citenamefont {Sirunyan}\ \emph
  {et~al.}(2018{\natexlab{a}})\citenamefont {Sirunyan} \emph
  {et~al.}}]{CMS:2017moi}%
  \BibitemOpen
  \bibfield  {author} {\bibinfo {author} {\bibfnamefont {A.~M.}\ \bibnamefont
  {Sirunyan}} \emph {et~al.} (\bibinfo {collaboration} {CMS}),\ }\href
  {\doibase 10.1007/JHEP03(2018)166} {\bibfield  {journal} {\bibinfo  {journal}
  {JHEP}\ }\textbf {\bibinfo {volume} {03}},\ \bibinfo {pages} {166} (\bibinfo
  {year} {2018}{\natexlab{a}})},\ \Eprint {http://arxiv.org/abs/1709.05406}
  {arXiv:1709.05406 [hep-ex]} \BibitemShut {NoStop}%
\bibitem [{\citenamefont {Aad}\ \emph {et~al.}(2021)\citenamefont {Aad} \emph
  {et~al.}}]{ATLAS:2021kog}%
  \BibitemOpen
  \bibfield  {author} {\bibinfo {author} {\bibfnamefont {G.}~\bibnamefont
  {Aad}} \emph {et~al.} (\bibinfo {collaboration} {ATLAS}),\ }\href {\doibase
  10.1007/JHEP07(2021)005} {\bibfield  {journal} {\bibinfo  {journal} {JHEP}\
  }\textbf {\bibinfo {volume} {07}},\ \bibinfo {pages} {005} (\bibinfo {year}
  {2021})},\ \Eprint {http://arxiv.org/abs/2103.01918} {arXiv:2103.01918
  [hep-ex]} \BibitemShut {NoStop}%
\bibitem [{\citenamefont {Buckley}\ \emph {et~al.}(2013)\citenamefont
  {Buckley}, \citenamefont {Butterworth}, \citenamefont {Grellscheid},
  \citenamefont {Hoeth}, \citenamefont {Lonnblad}, \citenamefont {Monk},
  \citenamefont {Schulz},\ and\ \citenamefont {Siegert}}]{Buckley:2010ar}%
  \BibitemOpen
  \bibfield  {author} {\bibinfo {author} {\bibfnamefont {A.}~\bibnamefont
  {Buckley}}, \bibinfo {author} {\bibfnamefont {J.}~\bibnamefont
  {Butterworth}}, \bibinfo {author} {\bibfnamefont {D.}~\bibnamefont
  {Grellscheid}}, \bibinfo {author} {\bibfnamefont {H.}~\bibnamefont {Hoeth}},
  \bibinfo {author} {\bibfnamefont {L.}~\bibnamefont {Lonnblad}}, \bibinfo
  {author} {\bibfnamefont {J.}~\bibnamefont {Monk}}, \bibinfo {author}
  {\bibfnamefont {H.}~\bibnamefont {Schulz}}, \ and\ \bibinfo {author}
  {\bibfnamefont {F.}~\bibnamefont {Siegert}},\ }\href {\doibase
  10.1016/j.cpc.2013.05.021} {\bibfield  {journal} {\bibinfo  {journal}
  {Comput. Phys. Commun.}\ }\textbf {\bibinfo {volume} {184}},\ \bibinfo
  {pages} {2803} (\bibinfo {year} {2013})},\ \Eprint
  {http://arxiv.org/abs/1003.0694} {arXiv:1003.0694 [hep-ph]} \BibitemShut
  {NoStop}%
\bibitem [{\citenamefont {Bierlich}\ \emph {et~al.}(2020)\citenamefont
  {Bierlich} \emph {et~al.}}]{Bierlich:2019rhm}%
  \BibitemOpen
  \bibfield  {author} {\bibinfo {author} {\bibfnamefont {C.}~\bibnamefont
  {Bierlich}} \emph {et~al.},\ }\href {\doibase 10.21468/SciPostPhys.8.2.026}
  {\bibfield  {journal} {\bibinfo  {journal} {SciPost Phys.}\ }\textbf
  {\bibinfo {volume} {8}},\ \bibinfo {pages} {026} (\bibinfo {year} {2020})},\
  \Eprint {http://arxiv.org/abs/1912.05451} {arXiv:1912.05451 [hep-ph]}
  \BibitemShut {NoStop}%
\bibitem [{\citenamefont {Buckley}\ \emph {et~al.}(2021)\citenamefont {Buckley}
  \emph {et~al.}}]{Buckley:2021neu}%
  \BibitemOpen
  \bibfield  {author} {\bibinfo {author} {\bibfnamefont {A.}~\bibnamefont
  {Buckley}} \emph {et~al.},\ }\href {\doibase
  10.21468/SciPostPhysCore.4.2.013} {\bibfield  {journal} {\bibinfo  {journal}
  {SciPost Phys. Core}\ }\textbf {\bibinfo {volume} {4}},\ \bibinfo {pages}
  {013} (\bibinfo {year} {2021})},\ \Eprint {http://arxiv.org/abs/2102.04377}
  {arXiv:2102.04377 [hep-ph]} \BibitemShut {NoStop}%
\bibitem [{\citenamefont {Alwall}\ \emph {et~al.}(2011)\citenamefont {Alwall},
  \citenamefont {Herquet}, \citenamefont {Maltoni}, \citenamefont {Mattelaer},\
  and\ \citenamefont {Stelzer}}]{Alwall:2011uj}%
  \BibitemOpen
  \bibfield  {author} {\bibinfo {author} {\bibfnamefont {J.}~\bibnamefont
  {Alwall}}, \bibinfo {author} {\bibfnamefont {M.}~\bibnamefont {Herquet}},
  \bibinfo {author} {\bibfnamefont {F.}~\bibnamefont {Maltoni}}, \bibinfo
  {author} {\bibfnamefont {O.}~\bibnamefont {Mattelaer}}, \ and\ \bibinfo
  {author} {\bibfnamefont {T.}~\bibnamefont {Stelzer}},\ }\href {\doibase
  10.1007/JHEP06(2011)128} {\bibfield  {journal} {\bibinfo  {journal} {JHEP}\
  }\textbf {\bibinfo {volume} {06}},\ \bibinfo {pages} {128} (\bibinfo {year}
  {2011})},\ \Eprint {http://arxiv.org/abs/1106.0522} {arXiv:1106.0522
  [hep-ph]} \BibitemShut {NoStop}%
\bibitem [{\citenamefont {Alwall}\ \emph {et~al.}(2014)\citenamefont {Alwall},
  \citenamefont {Frederix}, \citenamefont {Frixione}, \citenamefont {Hirschi},
  \citenamefont {Maltoni}, \citenamefont {Mattelaer}, \citenamefont {Shao},
  \citenamefont {Stelzer}, \citenamefont {Torrielli},\ and\ \citenamefont
  {Zaro}}]{Alwall:2014hca}%
  \BibitemOpen
  \bibfield  {author} {\bibinfo {author} {\bibfnamefont {J.}~\bibnamefont
  {Alwall}}, \bibinfo {author} {\bibfnamefont {R.}~\bibnamefont {Frederix}},
  \bibinfo {author} {\bibfnamefont {S.}~\bibnamefont {Frixione}}, \bibinfo
  {author} {\bibfnamefont {V.}~\bibnamefont {Hirschi}}, \bibinfo {author}
  {\bibfnamefont {F.}~\bibnamefont {Maltoni}}, \bibinfo {author} {\bibfnamefont
  {O.}~\bibnamefont {Mattelaer}}, \bibinfo {author} {\bibfnamefont {H.~S.}\
  \bibnamefont {Shao}}, \bibinfo {author} {\bibfnamefont {T.}~\bibnamefont
  {Stelzer}}, \bibinfo {author} {\bibfnamefont {P.}~\bibnamefont {Torrielli}},
  \ and\ \bibinfo {author} {\bibfnamefont {M.}~\bibnamefont {Zaro}},\ }\href
  {\doibase 10.1007/JHEP07(2014)079} {\bibfield  {journal} {\bibinfo  {journal}
  {JHEP}\ }\textbf {\bibinfo {volume} {07}},\ \bibinfo {pages} {079} (\bibinfo
  {year} {2014})},\ \Eprint {http://arxiv.org/abs/1405.0301} {arXiv:1405.0301
  [hep-ph]} \BibitemShut {NoStop}%
\bibitem [{\citenamefont {Sj\"ostrand}\ \emph {et~al.}(2015)\citenamefont
  {Sj\"ostrand}, \citenamefont {Ask}, \citenamefont {Christiansen},
  \citenamefont {Corke}, \citenamefont {Desai}, \citenamefont {Ilten},
  \citenamefont {Mrenna}, \citenamefont {Prestel}, \citenamefont {Rasmussen},\
  and\ \citenamefont {Skands}}]{Sjostrand:2014zea}%
  \BibitemOpen
  \bibfield  {author} {\bibinfo {author} {\bibfnamefont {T.}~\bibnamefont
  {Sj\"ostrand}}, \bibinfo {author} {\bibfnamefont {S.}~\bibnamefont {Ask}},
  \bibinfo {author} {\bibfnamefont {J.~R.}\ \bibnamefont {Christiansen}},
  \bibinfo {author} {\bibfnamefont {R.}~\bibnamefont {Corke}}, \bibinfo
  {author} {\bibfnamefont {N.}~\bibnamefont {Desai}}, \bibinfo {author}
  {\bibfnamefont {P.}~\bibnamefont {Ilten}}, \bibinfo {author} {\bibfnamefont
  {S.}~\bibnamefont {Mrenna}}, \bibinfo {author} {\bibfnamefont
  {S.}~\bibnamefont {Prestel}}, \bibinfo {author} {\bibfnamefont {C.~O.}\
  \bibnamefont {Rasmussen}}, \ and\ \bibinfo {author} {\bibfnamefont {P.~Z.}\
  \bibnamefont {Skands}},\ }\href {\doibase 10.1016/j.cpc.2015.01.024}
  {\bibfield  {journal} {\bibinfo  {journal} {Comput. Phys. Commun.}\ }\textbf
  {\bibinfo {volume} {191}},\ \bibinfo {pages} {159} (\bibinfo {year}
  {2015})},\ \Eprint {http://arxiv.org/abs/1410.3012} {arXiv:1410.3012
  [hep-ph]} \BibitemShut {NoStop}%
\bibitem [{\citenamefont {Dobbs}\ and\ \citenamefont
  {Hansen}(2000)}]{Dobbs:684090}%
  \BibitemOpen
  \bibfield  {author} {\bibinfo {author} {\bibfnamefont {M.}~\bibnamefont
  {Dobbs}}\ and\ \bibinfo {author} {\bibfnamefont {J.~B.}\ \bibnamefont
  {Hansen}},\ }\href {http://cds.cern.ch/record/684090} {\emph {\bibinfo
  {title} {{ATL-SOFT-2000-001}}}},\ \bibinfo {type} {Tech. Rep.}\ (\bibinfo
  {institution} {CERN},\ \bibinfo {address} {Geneva},\ \bibinfo {year}
  {2000})\BibitemShut {NoStop}%
\bibitem [{\citenamefont {Butterworth}\ \emph {et~al.}(2017)\citenamefont
  {Butterworth}, \citenamefont {Grellscheid}, \citenamefont {Kr\"amer},
  \citenamefont {Sarrazin},\ and\ \citenamefont
  {Yallup}}]{Butterworth:2016sqg}%
  \BibitemOpen
  \bibfield  {author} {\bibinfo {author} {\bibfnamefont {J.~M.}\ \bibnamefont
  {Butterworth}}, \bibinfo {author} {\bibfnamefont {D.}~\bibnamefont
  {Grellscheid}}, \bibinfo {author} {\bibfnamefont {M.}~\bibnamefont
  {Kr\"amer}}, \bibinfo {author} {\bibfnamefont {B.}~\bibnamefont {Sarrazin}},
  \ and\ \bibinfo {author} {\bibfnamefont {D.}~\bibnamefont {Yallup}},\ }\href
  {\doibase 10.1007/JHEP03(2017)078} {\bibfield  {journal} {\bibinfo  {journal}
  {JHEP}\ }\textbf {\bibinfo {volume} {03}},\ \bibinfo {pages} {078} (\bibinfo
  {year} {2017})},\ \Eprint {http://arxiv.org/abs/1606.05296} {arXiv:1606.05296
  [hep-ph]} \BibitemShut {NoStop}%
\bibitem [{\citenamefont {Korner}\ \emph {et~al.}(1993)\citenamefont {Korner},
  \citenamefont {Pilaftsis},\ and\ \citenamefont {Schilcher}}]{Korner:1992an}%
  \BibitemOpen
  \bibfield  {author} {\bibinfo {author} {\bibfnamefont {J.~G.}\ \bibnamefont
  {Korner}}, \bibinfo {author} {\bibfnamefont {A.}~\bibnamefont {Pilaftsis}}, \
  and\ \bibinfo {author} {\bibfnamefont {K.}~\bibnamefont {Schilcher}},\ }\href
  {\doibase 10.1016/0370-2693(93)91350-V} {\bibfield  {journal} {\bibinfo
  {journal} {Phys. Lett. B}\ }\textbf {\bibinfo {volume} {300}},\ \bibinfo
  {pages} {381} (\bibinfo {year} {1993})},\ \Eprint
  {http://arxiv.org/abs/hep-ph/9301290} {arXiv:hep-ph/9301290} \BibitemShut
  {NoStop}%
\bibitem [{\citenamefont {Herrero}\ \emph {et~al.}(2018)\citenamefont
  {Herrero}, \citenamefont {Marcano}, \citenamefont {Morales},\ and\
  \citenamefont {Szynkman}}]{Herrero:2018luu}%
  \BibitemOpen
  \bibfield  {author} {\bibinfo {author} {\bibfnamefont {M.~J.}\ \bibnamefont
  {Herrero}}, \bibinfo {author} {\bibfnamefont {X.}~\bibnamefont {Marcano}},
  \bibinfo {author} {\bibfnamefont {R.}~\bibnamefont {Morales}}, \ and\
  \bibinfo {author} {\bibfnamefont {A.}~\bibnamefont {Szynkman}},\ }\href
  {\doibase 10.1140/epjc/s10052-018-6281-3} {\bibfield  {journal} {\bibinfo
  {journal} {Eur. Phys. J. C}\ }\textbf {\bibinfo {volume} {78}},\ \bibinfo
  {pages} {815} (\bibinfo {year} {2018})},\ \Eprint
  {http://arxiv.org/abs/1807.01698} {arXiv:1807.01698 [hep-ph]} \BibitemShut
  {NoStop}%
\bibitem [{\citenamefont {Brdar}\ \emph {et~al.}(2020)\citenamefont {Brdar},
  \citenamefont {Lindner}, \citenamefont {Vogl},\ and\ \citenamefont
  {Xu}}]{Brdar:2020nbj}%
  \BibitemOpen
  \bibfield  {author} {\bibinfo {author} {\bibfnamefont {V.}~\bibnamefont
  {Brdar}}, \bibinfo {author} {\bibfnamefont {M.}~\bibnamefont {Lindner}},
  \bibinfo {author} {\bibfnamefont {S.}~\bibnamefont {Vogl}}, \ and\ \bibinfo
  {author} {\bibfnamefont {X.-J.}\ \bibnamefont {Xu}},\ }\href {\doibase
  10.1103/PhysRevD.101.115001} {\bibfield  {journal} {\bibinfo  {journal}
  {Phys. Rev. D}\ }\textbf {\bibinfo {volume} {101}},\ \bibinfo {pages}
  {115001} (\bibinfo {year} {2020})},\ \Eprint
  {http://arxiv.org/abs/2003.05339} {arXiv:2003.05339 [hep-ph]} \BibitemShut
  {NoStop}%
\bibitem [{\citenamefont {Patel}(2015)}]{Patel:2015tea}%
  \BibitemOpen
  \bibfield  {author} {\bibinfo {author} {\bibfnamefont {H.~H.}\ \bibnamefont
  {Patel}},\ }\href {\doibase 10.1016/j.cpc.2015.08.017} {\bibfield  {journal}
  {\bibinfo  {journal} {Comput. Phys. Commun.}\ }\textbf {\bibinfo {volume}
  {197}},\ \bibinfo {pages} {276} (\bibinfo {year} {2015})},\ \Eprint
  {http://arxiv.org/abs/1503.01469} {arXiv:1503.01469 [hep-ph]} \BibitemShut
  {NoStop}%
\bibitem [{\citenamefont {Hahn}\ and\ \citenamefont
  {Perez-Victoria}(1999)}]{Hahn:1998yk}%
  \BibitemOpen
  \bibfield  {author} {\bibinfo {author} {\bibfnamefont {T.}~\bibnamefont
  {Hahn}}\ and\ \bibinfo {author} {\bibfnamefont {M.}~\bibnamefont
  {Perez-Victoria}},\ }\href {\doibase 10.1016/S0010-4655(98)00173-8}
  {\bibfield  {journal} {\bibinfo  {journal} {Comput. Phys. Commun.}\ }\textbf
  {\bibinfo {volume} {118}},\ \bibinfo {pages} {153} (\bibinfo {year}
  {1999})},\ \Eprint {http://arxiv.org/abs/hep-ph/9807565}
  {arXiv:hep-ph/9807565} \BibitemShut {NoStop}%
\bibitem [{\citenamefont {Sirunyan}\ \emph
  {et~al.}(2018{\natexlab{b}})\citenamefont {Sirunyan} \emph
  {et~al.}}]{CMS:2017con}%
  \BibitemOpen
  \bibfield  {author} {\bibinfo {author} {\bibfnamefont {A.~M.}\ \bibnamefont
  {Sirunyan}} \emph {et~al.} (\bibinfo {collaboration} {CMS}),\ }\href
  {\doibase 10.1007/JHEP06(2018)001} {\bibfield  {journal} {\bibinfo  {journal}
  {JHEP}\ }\textbf {\bibinfo {volume} {06}},\ \bibinfo {pages} {001} (\bibinfo
  {year} {2018}{\natexlab{b}})},\ \Eprint {http://arxiv.org/abs/1712.07173}
  {arXiv:1712.07173 [hep-ex]} \BibitemShut {NoStop}%
\bibitem [{\citenamefont {Bambade}\ \emph {et~al.}(2019)\citenamefont {Bambade}
  \emph {et~al.}}]{Bambade:2019fyw}%
  \BibitemOpen
  \bibfield  {author} {\bibinfo {author} {\bibfnamefont {P.}~\bibnamefont
  {Bambade}} \emph {et~al.},\ }\href@noop {} {\  (\bibinfo {year} {2019})},\
  \Eprint {http://arxiv.org/abs/1903.01629} {arXiv:1903.01629 [hep-ex]}
  \BibitemShut {NoStop}%
\bibitem [{\citenamefont {Dong}\ \emph {et~al.}(2018)\citenamefont {Dong} \emph
  {et~al.}}]{CEPCStudyGroup:2018ghi}%
  \BibitemOpen
  \bibfield  {author} {\bibinfo {author} {\bibfnamefont {M.}~\bibnamefont
  {Dong}} \emph {et~al.} (\bibinfo {collaboration} {CEPC Study Group}),\
  }\href@noop {} {\  (\bibinfo {year} {2018})},\ \Eprint
  {http://arxiv.org/abs/1811.10545} {arXiv:1811.10545 [hep-ex]} \BibitemShut
  {NoStop}%
\bibitem [{\citenamefont {Abada}\ \emph {et~al.}(2019)\citenamefont {Abada}
  \emph {et~al.}}]{FCC:2018evy}%
  \BibitemOpen
  \bibfield  {author} {\bibinfo {author} {\bibfnamefont {A.}~\bibnamefont
  {Abada}} \emph {et~al.} (\bibinfo {collaboration} {FCC}),\ }\href {\doibase
  10.1140/epjst/e2019-900045-4} {\bibfield  {journal} {\bibinfo  {journal}
  {Eur. Phys. J. ST}\ }\textbf {\bibinfo {volume} {228}},\ \bibinfo {pages}
  {261} (\bibinfo {year} {2019})}\BibitemShut {NoStop}%
\bibitem [{\citenamefont {Calibbi}\ \emph {et~al.}(2021)\citenamefont
  {Calibbi}, \citenamefont {Marcano},\ and\ \citenamefont
  {Roy}}]{Calibbi:2021pyh}%
  \BibitemOpen
  \bibfield  {author} {\bibinfo {author} {\bibfnamefont {L.}~\bibnamefont
  {Calibbi}}, \bibinfo {author} {\bibfnamefont {X.}~\bibnamefont {Marcano}}, \
  and\ \bibinfo {author} {\bibfnamefont {J.}~\bibnamefont {Roy}},\ }\href@noop
  {} {\  (\bibinfo {year} {2021})},\ \Eprint {http://arxiv.org/abs/2107.10273}
  {arXiv:2107.10273 [hep-ph]} \BibitemShut {NoStop}%
\end{thebibliography}
%\end{document}
%\include{bibcom}
%
\end{document}